\documentclass[fleqn,usenatbib]{arxiv_mnras}
\usepackage{newtxtext,newtxmath} 
\usepackage[whole]{bxcjkjatype}
\usepackage[T1]{fontenc} 
\DeclareRobustCommand{\VAN}[3]{#2}
\let\VANthebibliography\thebibliography
\def\thebibliography{\DeclareRobustCommand{\VAN}[3]{##3}\VANthebibliography}
\usepackage{graphicx}	
\usepackage{amsmath}	
\usepackage{bm}  
\usepackage{color}   
\usepackage{mfirstuc}
      
\def\deg{^\circ}    \def\be{\begin{equation}} \def\ee{\end{equation}}     \def\kms{km s$^{-1}$}
  
\def\Alf{Alfv\'en } \def\alf{Alfv\'en }  \def\Va{V} \def\va{$V$}        
 \def\tu{t_{\rm unit}}\def\pr{p_r}\def\d{\partial} \def\pth{p_\theta}\def\pphi{p_\phi}\def\cot{{\rm cot}\ } \def\sin{{\rm sin}\ } \def\cos{ {\rm cos}} \def\f{\frac}   
 \def\ss{\subsection}\def\sss{\subsubsection}

\title[MHD lensing in ISM and SNR morphology]{MHD lensing in inhomogeneous ISM for qualitative understanding of  the morphology of supernova remnants}

 \author[Y. Sofue]{Yoshiaki Sofue \\ 
Institute of Astronomy, The University of Tokyo, Mitaka, Tokyo 186-0015, Japan}

\date{Accepted for publication in Astrophysics \& Space Sciences (ApSS)} 
\pubyear{2024}  

\begin{document} 
\maketitle      

\begin{abstract}
Morphological evolution of expanding shells of fast-mode magnetohydrodynamic (MHD) waves through an inhomogeneous ISM is investigated in order to qualitatively understand the complicated morphology of shell-type supernova remnants (SNR).  
Interstellar clouds with high \alf velocity act as concave lenses to diverge the MHD waves, while those with slow \alf velocity act as convex lenses to converge the waves to the focal points. 
By combination of various types of clouds and fluctuations with different \alf velocities, sizes, or wavelengths, the MHD-wave shells attain various morphological structures, exhibiting filaments, arcs, loops, holes, and focal strings, mimicking old and deformed SNRs.  
\end{abstract}


\section{Introduction}
 
The interaction of expanding blast waves of supernova remnants (SNR) with the interstellar medium is one of the most fundamental subjects of interstellar physics 
\citep{1975ApJ...195..715M,
1985A&A...145...70T,
1992ApJ...390L..17S,
1999ApJ...524..192S,1999ApJ...511..798C}.
The shock compression of gas clouds and magnetic fields by SNRs is also an important issue in the high-energy astrophysics for investigating the acceleration mechanisms of cosmic-rays. 
Extensive observations of the interacting clouds near SNRs have been obtained from $\gamma$-ray to radio wavelengths, on which several models for cosmic-ray accelerations have been proposed
\citep{2009ApJ...707L.179F,
2009ApJ...698L.133A,
2012ApJ...744...71I,
2014A&A...565A..74C,
2017ApJ...836...23A,
2019ApJ...876...37S,
2020ApJ...902...53S,
2020ApJ...894...51A,
2022NatCo..13.5098G,
2023PASJ...75..338S}. 

Multi-wavelengths observations from $\gamma$ to radio have been obtained also for investigating the morphological evolution of SNRs interacting with the ISM, which are used at the same time to diagnose the physical conditions and magnetic fields of the ISM
\citep{2000A&A...359..316B,
2011A&A...531A.129B,
2011ApJ...732..114L,
2006A&A...458..213M, 
2017A&A...599A..45M,
2008ApJ...676.1050B,
2016A&A...587A.148W,
2021ApJ...919..123S, 
2022A&A...666A...2O,
2021ApJ...910..149O}. 
In order to explain the observed properties and morphologies of SNRs, a number of numerical simulations of hydro- and magneto-hydrdoyanamics have been obtained about the interaction of shock waves with the inhomogeneous/turbulent ISM and stellar winds
\citep{2002AJ....124.2145V,
2003ApJ...584..284V,
2006MNRAS.371..369S,
2014MNRAS.445.2484F,
2014MNRAS.442..229T,
2017MNRAS.472.2117M,
2019MNRAS.482.1602Z,
2021A&A...649A..14U,
2022MNRAS.513.3345K}.

The current studies of the SNR-cloud interaction have been devoted mainly to study the effects of SNRs as the active source on the shocked medium to change the physical conditions interior to the SNR shell. 
The interaction of an SNR with the ISM takes place most directly at the contact surface of the outermost blast wave, which is observed as a thin shell with optical emission due to the recombination process between the blast and cold ISM as well as the non-thermal radio emission due to the Fermi acceleration of cosmic-rays at the highest compression surface.

At the same time, the morphology of SNRs also manifests the past paths of propagation of the shell front through the interstellar medium, because the blast wave propagation is strongly influenced by the density and magnetic fields in the ISM.
This means that the morphology of SNRs may be used to diagnose the gas density and magnetic field distributions in the ISM before the passage of SNR shells \citep{1978A&A....67..409S}.
Therefore, morphological study of images of aged SNRs in optical  \citep{1973ApJS...26...19V}
and radio-continuum emissions
\citep{2014PASA...31...42G,
2006A&A...457.1081K,
2006AJ....131.2525H}, 
would provide with a tool to diagnose the interstellar medium by analysing the shapes that have been deformed during the propagation. 

In this paper we aim at understanding the morphology of old shell-type SNRs by the deformation of expanding shells in the inhomogeneous ISM.
We show that various types of morphology of old SNRs are qualitatively understood by the lensing effect due to the variation of propagation speed of the waves in the ISM.
Since the treatment of the wave in this paper is linear, we discuss only the geometrical shapes of the front, or the morphology, of old SNRs.
Non-linear physical quantities such as the density, magnetic strength, cosmic-ray acceleration, emission rates, etc.. cannot be calculated in the present linear approximation.  
However, the relative distribution of the brightness of wave front, which defines the morphology, can be mapped by the distribution of column density of wave packets projected on the sky. 
Thereby, the brightness of a sheet of MHD wave is highest, when it is seen edge on and the tangential sheet shines brightest, whereas a face-on sheet is almost invisible. 

In section \ref{sec2} we formulate the method using fast-mode MHD waves. In section \ref{sec3} we describe result of simulation for  large-scale deformation of shell fronts. Section \ref{sec3} deals with more complicated morphology, and section \ref{sec4} compares the model with the well-known SNRs having unique morphology.
In section \ref{secdiscussion} we present some speculative considerations towards understanding the morphology of SNRs in more general cases based on the MHD model, and discuss the difficulty and limitation of the model. Section \ref{secsummary} summarizes the results. 

\section{MHD lensing}  
\label{sec2}

\ss{Fast-mode MHD wave and \alf velocity in the ISM}

The \alf velocity in a medium with magnetic-field strength $B$ and gas density $\rho$ is given by
\be
V=B/\sqrt{4 \pi \rho}.
\ee
Its variation with the cloud density depends on the coupling property of the magnetic field and gas.
There are three types of coupling:
\begin{itemize}
    \item (a) Frozen-in field depends on the gas density as $B\propto \rho^{2/3}$, which leads to \alf velocity varying as $\Va \propto \rho^{1/6}$. This happens when the magnetized gas cloud is highly turbulent.
    \item (b) Cylindrical magnetic field uniformly filled with the gas varies as $B\propto \rho$, so that we have $V\propto \rho^{1/2}$. This occurs in stretched clouds and filaments extending along the magnetic lines of force.
    \item (c) Gas flowing along magnetic lines of force with the field strength kept consant yields variation of the \alf velocity as $V\propto \rho^{-1/2}$. 
    Such a flow is found in a magnetic tube inflating from the Galactic plane by the Parker instability with a condensed cloud at the bottom of the tube.
\end{itemize}
Interstellar clouds in cases (a) and (b) yield higher \va than the ambient medium, so that they act as concave lenses to diverge the MHD waves.
Clouds in case (c) act as convex lenses to converge the waves.
In this paper we represent the property of a cloud by its \va variation relative to the mean value, and represent the variation by equation \ref{eqV}.

MHD disturbances excited by explosive events such as due to SNe in the interstellar space propagate as spherical shock waves in the initial expansion phase. 
In the fully expanded phase {when the expanding velocity becomes comparable to the local \alf velocity or sound velocity, they propagate as the fast-mode MHD compression wave (hereafter, MHD wave), Alfv\'en wave, and the sound wave.  }

The typical \alf velocity is $\Va\sim 10$ \kms with magnetic strength $B\sim 5~ \mu$G and gas density $\rho\sim 1$ H cm$^{-3}$, whereas the sound velocity is $c_{\rm s}\sim 0.1$ and 1.3 \kms for molecular and HI gases with temperatures $T_{\rm mol}\sim 20$ K and $T_{\rm HI}\sim 100$ K, respectively.
Therefore, among the three modes, the fast MHD wave is most efficient to convey the energy into the ISM, which propagates at the group velocity approximately equal to the Alfv\'en velocity, $v_{\rm g}\simeq \Va$, where the \alf velocity is higher than the sound velocity, or the magnetic pressure is higher than the thermal pressure \citep{1994ApJ...435..482P}. 

   \ss{{\alf vs turbulent velocities}}
    
One concern that disturbs the above assumption is that the \alf velocity is comparable to the turbulent velocity as the consequence of the energy balance (equilibrium) between the magnetic and thermal energy densities in the Galactic disc.
Because the Eikonal equations hold locally, the wave propagation is superposed by the local turbulent motion.
This effect is significant on the ray paths slowly bent (refracted) by low-\va clouds.
On the other hand, it does not affect so much the ray paths against strongly disturbed (reflected) ray paths by high-\va lensing.
In this paper, we assume that the turbulent motion is negligible, which, even if existed, disturbs the result on the same order as that due to low-\va lensing.
The amount of ray deflection is much smaller than that by high-\va lensing, and the entire morphological shape is more effectively determined by the high-\va lensing.

\ss{Propagation of MHD waves by Eikonal equations}

The propagation of MHD waves is traced by solving the Eikonal equations developed for the Solar coronal Moreton waves  \citep{1970PASJ...22..341U,1974SoPh...39..431U,2024NatCo..15.3281Z}.
The method has been applied to the SNRs  \citep{1978A&A....67..409S,2023MNRAS.518.6273S} as well as to the Galactic Centre explosions \citep{1978A&A....67..409S,2023MNRAS.518.6273S}.
{Gas clouds with higher density with constant and weaker magnetic strength act as a convex lens (low-\va lens) gathering the waves.
On the other hand, gas clouds with frozen-in magnetic field and gaseous cavity (hole) with constant and higher magnetic strength act as a concave lens (high-\va lens) reflecting the waves.}

\begin{figure} 
\begin{center}      
(A) Low-\va cloud (convex lens)  \\ 
\includegraphics[width=6.9cm]{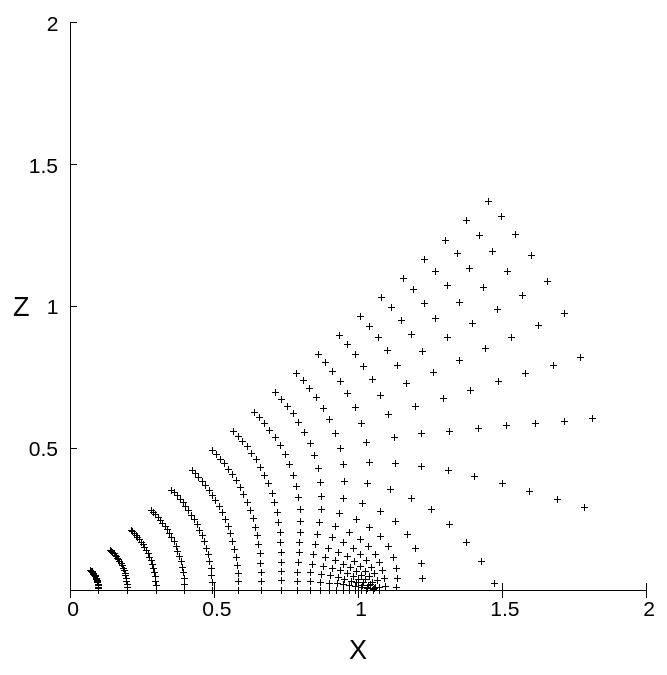}  \\
(B) High-\va cloud (concave lens) \\
\includegraphics[width=6.9cm]{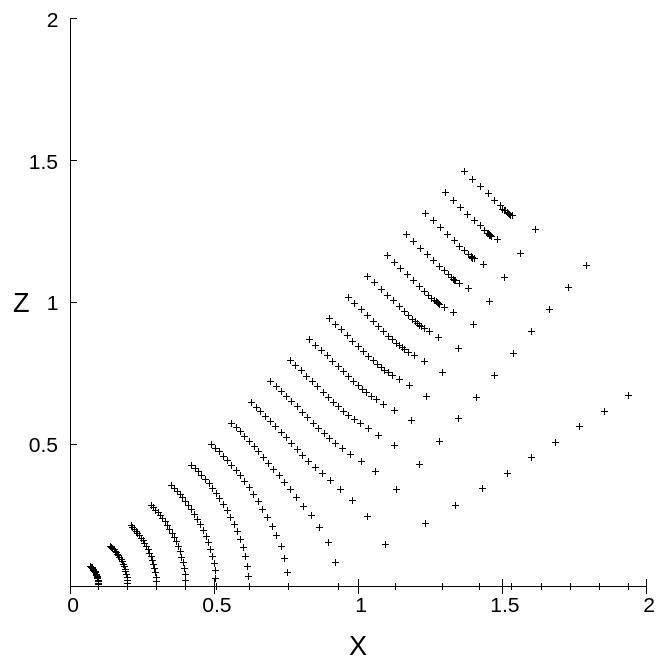} 
\end{center}   
\caption{MHD ray path from an explosion center encountered by spherical clouds located at $x=1$ with (A) lower and (B) higher \va, acting as convex and concave lens, respectively \citep{1978A&A....67..409S}.} 
\label{raypath} 
\end{figure}         

The Eikonal equations describing the propagation of an MHD wave packet are given as follows \citep{1970PASJ...22..341U,1974SoPh...39..431U}.
\be \f{dr}{dt}=V \f{\pr}{p}, \ee
\be \f{d\theta}{dt}=V \f{\pth}{rp}, \ee 
\be \f{d\phi}{dt}=V \f{\pphi}{rp\  \sin \theta}, \ee 
\be \f{d\pr}{dt}=-p\f{\d V}{\d r} +\f{V}{rp}(\pth^2+\pphi^2), \ee
\be \f{d\pth}{dt}=-\f{p}{r} \f{\d V}{\d \theta} 
- \f{V}{rp}(\pth \pr-\pphi^2 \ \cot \ \theta), \ee 
\be \f{d\pphi}{dt}=-\f{p}{\sin \theta} \f{\d V}{\d \phi} - \f{V}{rp}(\pphi \pr + \pphi \pth \ \cot \ \theta), \ee 
where
$V=V(r,\theta,\phi)=V(x,y,z)$ is the \Alf velocity, the vector ${\bf \it p}=(\pr, \pth, \pphi)={\rm grad}\  \Phi$ is defined by the gradient of the eikonal $\Phi$, $p=(\pr^2+\pphi^2+\pth^2)^{1/2}$, $(r,\theta, \phi)$ and $(x,y,z)$ are the polar and Cartesian coordinates.   

An interstellar cloud with lower \alf velocity deflects the path toward its center, acting as a convex lens, and the converging waves are focused around its focal point to produce amplified wave front.
On the other hand, a cloud with higher \alf velocity acts as a concave lens, and diverges the waves to produce a hole with the deflected waves making a ring.

Figure \ref{raypath} shows the propagation of MHD wave packets produced at the center intervened by clouds with higher and lower \alf velocities than the ambient velocity, where the \va variation is expressed by Gaussian function with an $e$-folding scale radius of 0.5.

In the following simulations, we solve the Eikonal equations assuming the functional form of the \va variation in the space, which is expressed in the following form:
\be
V=V_0[1+\epsilon \ f(x,y,z)],
\label{eqV}
\ee
where $V_0$ is the mean \va in the ISM, $f(x,y,z)$ is a function expressing the shape of the turbulence, and $\epsilon$ represents the relative amplitude of variation.
Here, $\epsilon \ge -1$, and low- and high-\va clouds have negative and positive $\epsilon$, respectively. 

\section{{Large-scale deformation}}
\label{sec3}

\ss{Lopsidedness by \va gradient} 

\begin{figure*} 
\begin{center}      
(A) \va gradient \hskip 2cm (B) Low-\va disc + high-\va halo \hskip 2cm 
\\
\includegraphics[width=5.6cm]{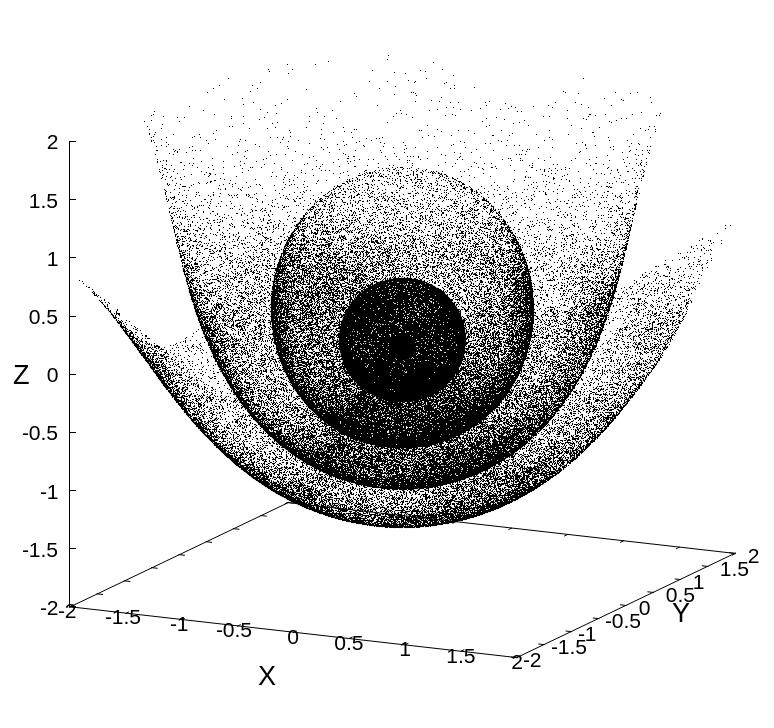} 
\includegraphics[width=5.6cm]{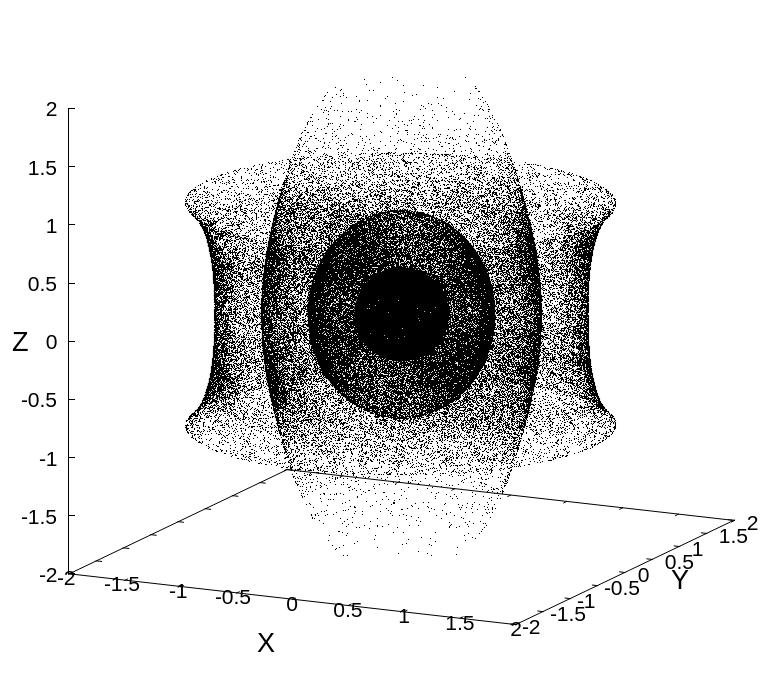}  
\\
(C) Thin low-\va Gaussian disc \hskip 2cm (D) High-\va Gaussian disc\\
\includegraphics[width=6cm]{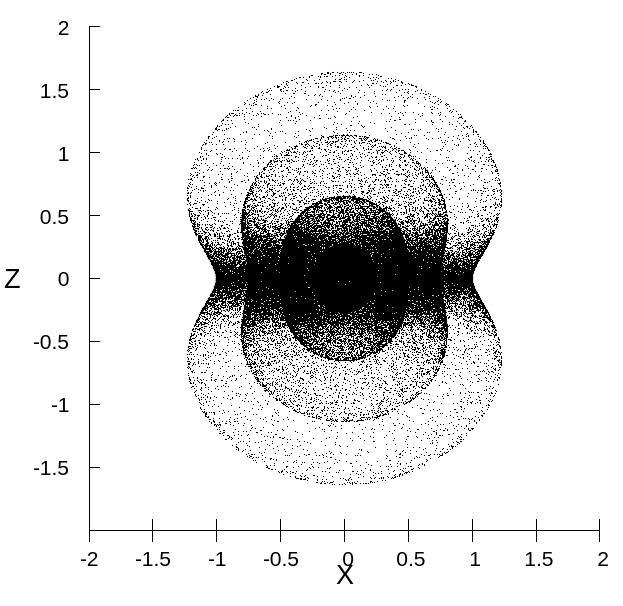}  
\includegraphics[width=6cm]{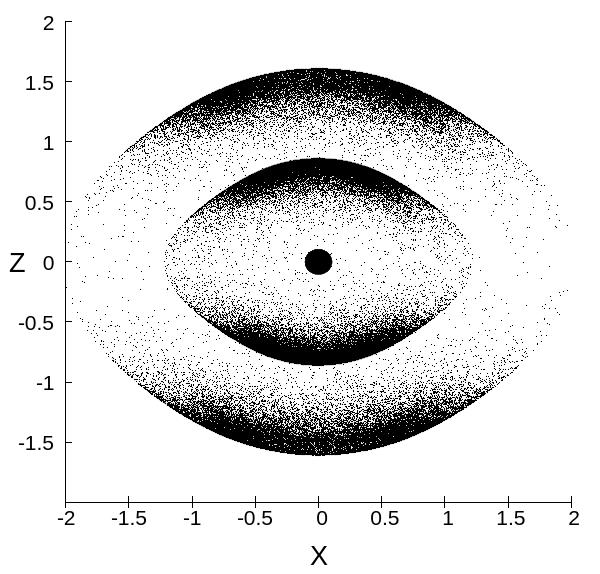}  
\end{center}   
\caption{{Explosion in a medium having (A) exponential-gradient in $z$ direction, in a (B) hyperbolic-cosine $z$ disc, (C)  flat negative Gaussian disc (low-\va disc), and (D) flat positive Gaussian disc (high-\va disc). } }
\label{disc} 
\end{figure*}

The gradient of \alf velocity produces a lopsided elongation of the front shape, so that the lower-\va side is stopped, and the shell expands faster in the higher-\va side.
Figure \ref{disc} (A) shows a case when the {wave expansion takes} place in an ISM with one-directional \va variation expressed by 
\be
{V =V_0 \ [1+\exp (z/h)],}
\ee
where $V_0$ and $h=0.5$ are constants. 
Hereafter, the coordinates are $(x,y,z)$ measured in arbitrary unit.
The time is measured in unit of unity scale divided by \va, {$\tu=1/V_0$}.
In this diagram, the wave front at $t=0.25, 0.5, 0.75$ and 1 $\tu$ are plotted.
The SN explosion takes place at $(x,y,z)=(0,0,0)$ spherically, and the wave front is spherical in the initial stage.
As it expands, the front is elongated in the $z$ direction of increasing \va, and attains an open cone structure.
The front finally forms a one-sided curved sheet open toward positive higher-\va direction.

\def\cosh{ {\rm cosh} }

\begin{figure*} 
\begin{center}      
(A) High-\va plate \hskip 5cm (B) Low-\va plate\\
\includegraphics[width=6cm]{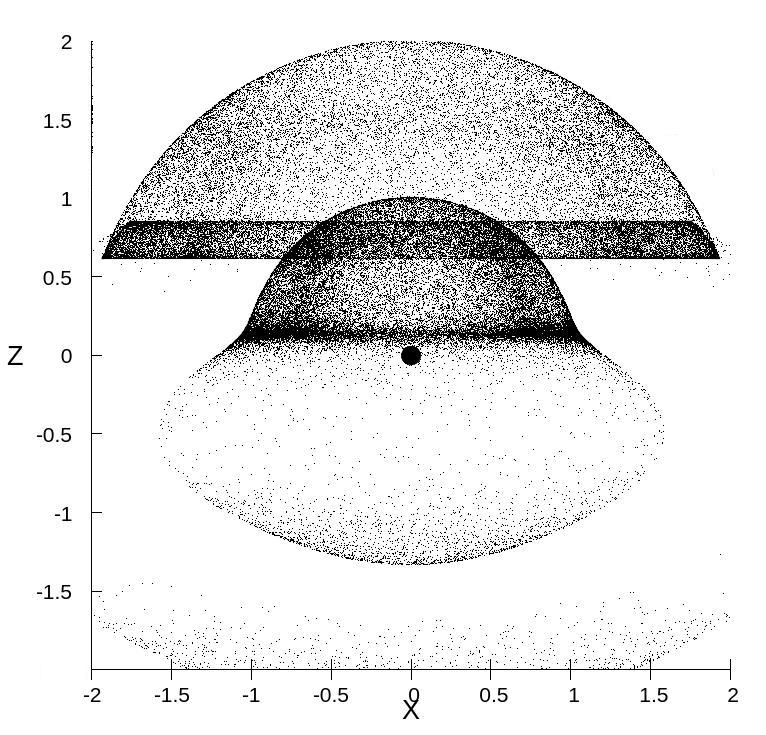} 
\includegraphics[width=6cm]{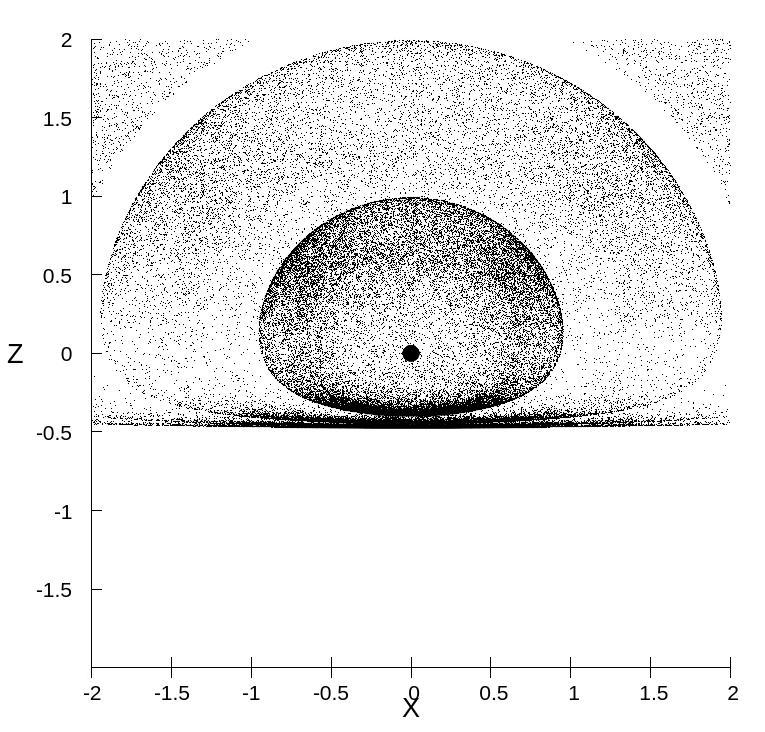}   
\end{center}
\caption{Encounter with widely extended (A) high- and (B) low-\va plate (wall). 
} 
\label{wall}  
\end{figure*}

If the {wave expansion} takes place in a disc with increasing \va gradient toward both sides expressed by a cosh function,
\be
{V=V_0 [1+0.5 \cosh(z/0.5)]},
\ee
the front shape becomes bi-conical, as shown in figure \ref{disc} (B) for $z_0=0$. 
If the disc is thin enough compared with the expanding shell radius, the front becomes bipolar, exhibiting a dumbbell-shaped bubbles for negative-Gaussian disc (figure \ref{wall} (C)). 
If the disc has higher-\va than the surrounding ISM having positive-Gaussian distribution, the front attains bipolar,or bilateral caps (figure \ref{disc} (D)), somehow mimicking such an axisymmetric bilateral SNR as G296.5+10.0 \citep{2000AJ....119..281G}.
 
\ss{High- and low-\va walls}

If the {MHD-wave front} encounters a wall with sufficiently high \va with finite thickness, the front is reflected and the shell is deformed to become a half-spherical shape.
Figure \ref{wall} (A) shows an example of such an encounter, where a horizontal wall is located at $z_0=0.5$ expressed by
\be
V=V_0 [1+\exp ((z-z_0)/h)^2]
\ee
with $z=0.5$.

If the wall has lower \va, the front is strongly refracted toward the wall and absorbed there, resulting in a half shell, and dense, flat front along the wall is formed, as shown in panel (B) of figure \ref{wall}.

\ss{Clouds with low-\va vs high-\va: Convex vs concave lenses}

\def\exp{{\rm exp} }

\begin{figure*}   
	\begin{center}
(A) Low-\va cloud \hskip 5cm (B) High-\va cloud \\
\includegraphics[width=7cm]{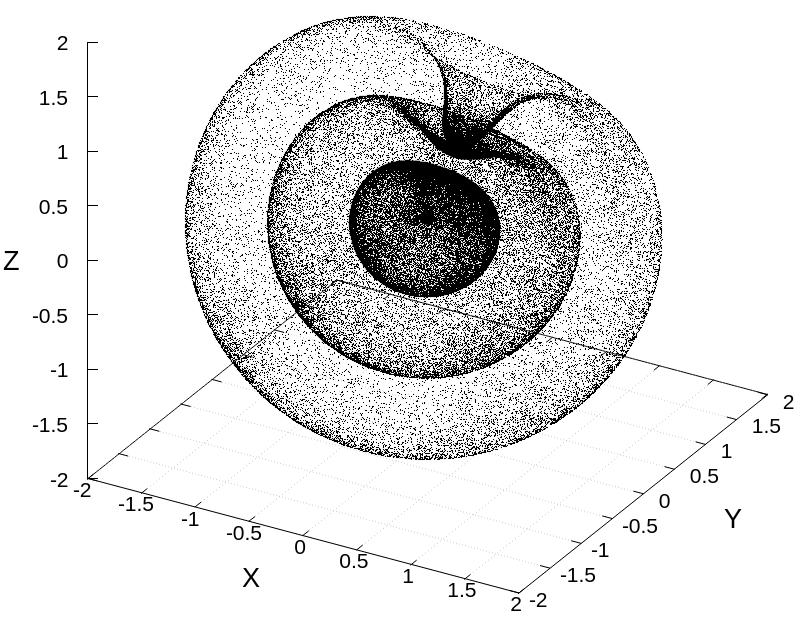}   
\includegraphics[width=6.5cm]{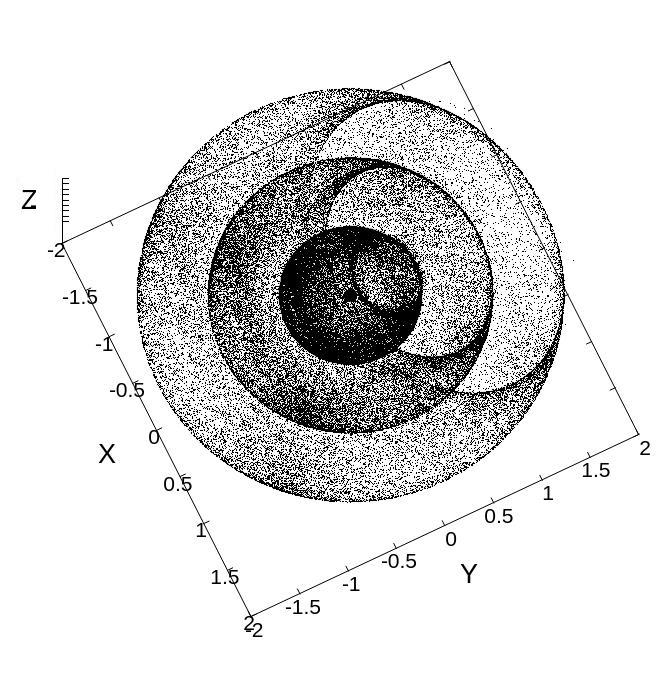}   
\end{center} 
\caption{{(A) Encounter with a low \va cloud. 
) 
(B) Encounter with a high-V cloud, making a hole surrounded by a ring of amplified wave density. 
}}
\label{Vcloud} 
	\end{figure*}

{A low-\va cloud acts as a convex lens for the ray paths of the MHD wave packets.}
Figure \ref{Vcloud} shows front propagation of a spherical MHD wave front encountering a low-\va cloud, where the \alf velocity is expressed by a negative Gaussian function as
\be
V=V_0[1+\epsilon \ \exp(-(r/r_0)^2)].
\ee
Here, $r=[(x-x_0)^2+(y-y_0)^2+(z-z_0)^2]^{1/2}$ is the distance from the cloud center at $(x,y,z)=(x_0,y_0,z_0)\sim (0.5,0,5,0,5)$, $r_0\sim 1$ is the radius of the cloud, and $\epsilon\sim -0.5$ is the negative amplitude of \va.
 
The front is reflected by the lens and bent toward the focal point of the cloud.
Accordingly, the wave forms a concave surface shape and the amplitude increases. 

If \va of the cloud is small enough, the front evolves into an elephant-trunk shaped elongated lobe. 
However, the focusing effect to the trunk's head amplifies the magnetic field together with wave density, which works to increase the \alf velocity and suppresses the growth of the trunk.
 
On the other hand, when the \alf velocity of the cloud is higher than the ambient medium, it acts as a concave lens.
The ray paths are diverged by the lens, and the front expands locally, making a vacant hole.
Thereby, reflected rays are locally converged on the edge of the hole, and form a focal ring (loop) of amplified waves.
Figure \ref{Vcloud} shows an example of evolution of the front through such a high-\va cloud with $\epsilon\sim +0.5$.

\ss{Large-scale deformation by large clouds}

\begin{figure*} 
\begin{center}      
(A) high-\va plate: IC433\\
\includegraphics[width=7.8cm]{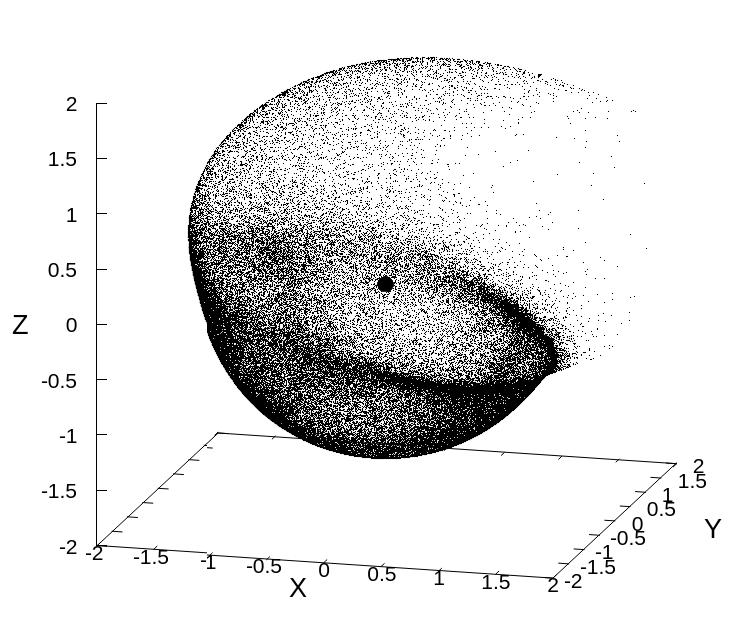}   
\includegraphics[width=6.cm]{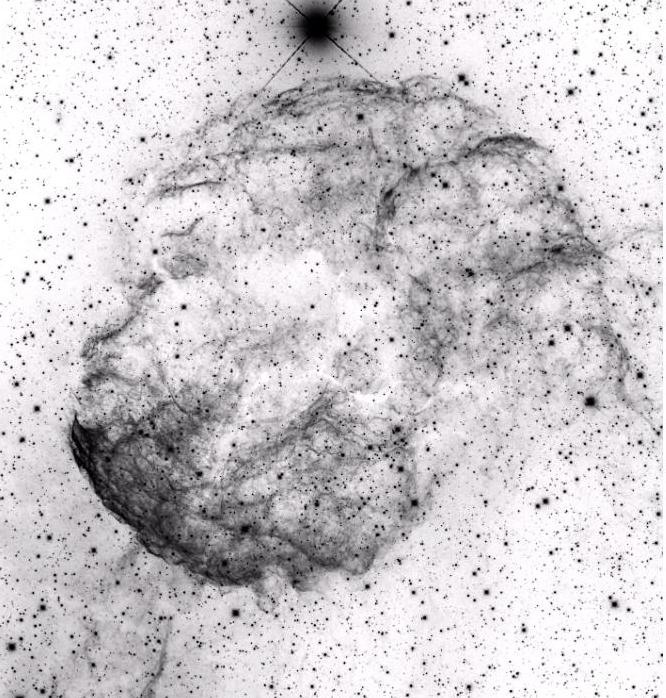}  \\
(B) High-\va plate: Sh2\\ 
\includegraphics[width=7.8cm]{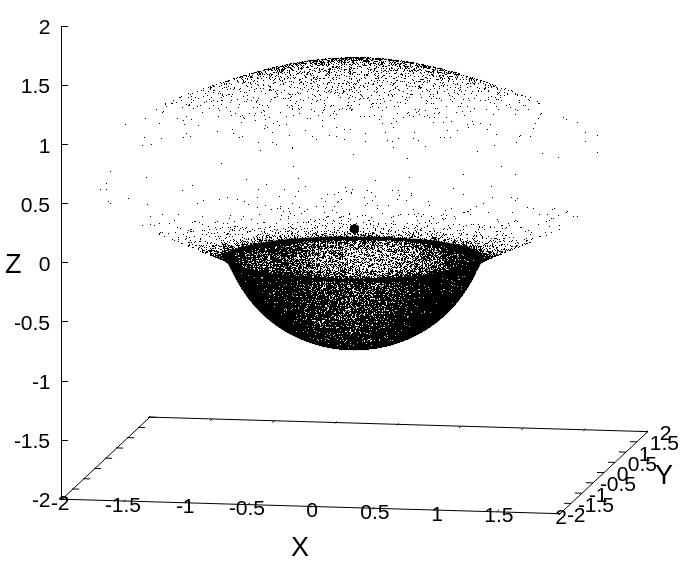}   
\includegraphics[width=6.7cm]{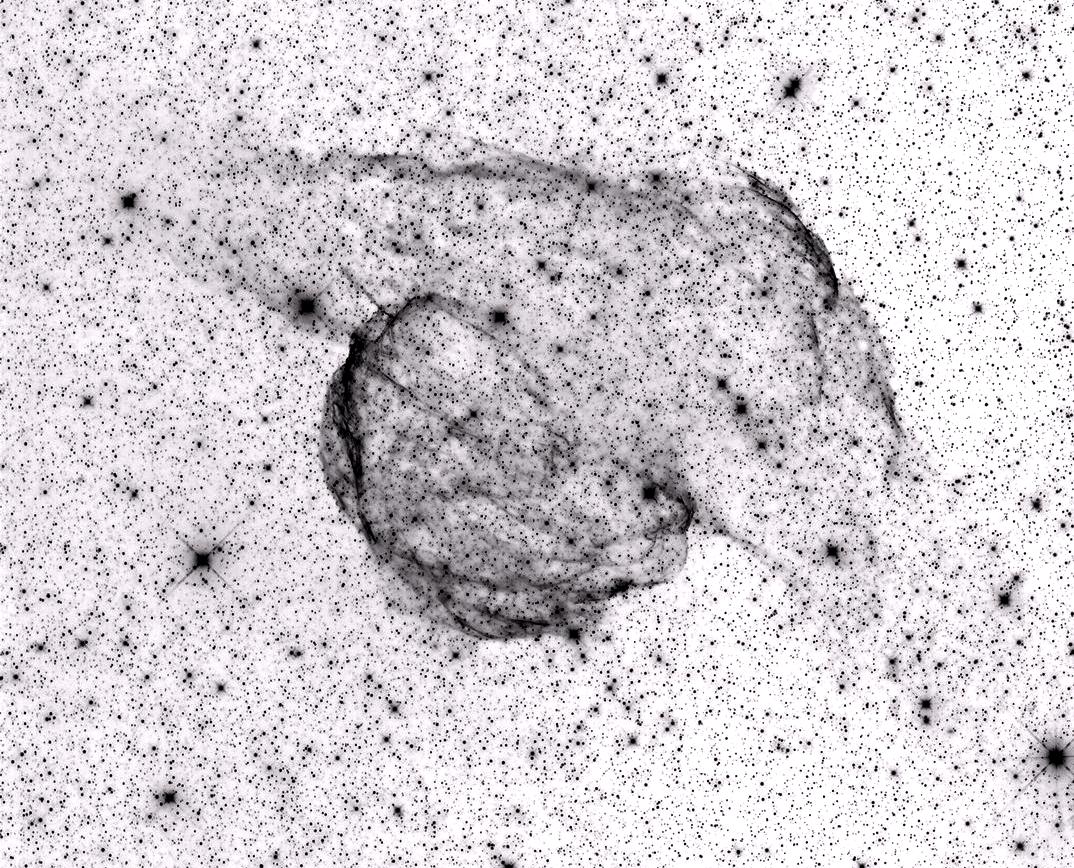}  
\end{center}
\caption{(A) Dual half shells with different sizes divided by the encounter with a high-\va plate, mimicking the SNR IC 443 (
https://apod.nasa.gov/apod/ap231226.html) 
(B) Dual half shells divided by a high-\va plate simulating SNR Sh 2-224 (wing-and-shell nebula, VRO 42.05.01) \citep{2024A&A...684A.178A} 
(https://apod.astronomia.com/wp-content/uploads/2022/01/sh2-224.jpg).
} 
\label{ic443} 
\end{figure*}

\sss{IC443 and Sh 2}
Large-scale deformation such as observed for the SNRs IC 443  (figure \ref{ic443}) can be 
explained by interaction of a larger-sized clouds and density fluctuations than the size of the expanding front.
When a spherical front encounters a wall or a sheet with higher-\va, the front suddenly expands faster and the shape becomes lopsided, so that one side of the shell is more extended.
{
Our MHD-wave simulation of IC 443 is shown in figure \ref{ic443}, where the large-scale deformation of the shell is reasonably mimicked in coincidence with that obtained by the 3D numerical simulations \citep{2021A&A...649A..14U}.
}

\ss{Manatee (W50) SNR}

\begin{figure*} 
\begin{center}       
(A) Manatee: Explosion in a high-\va tube\\
\includegraphics[width=8cm]{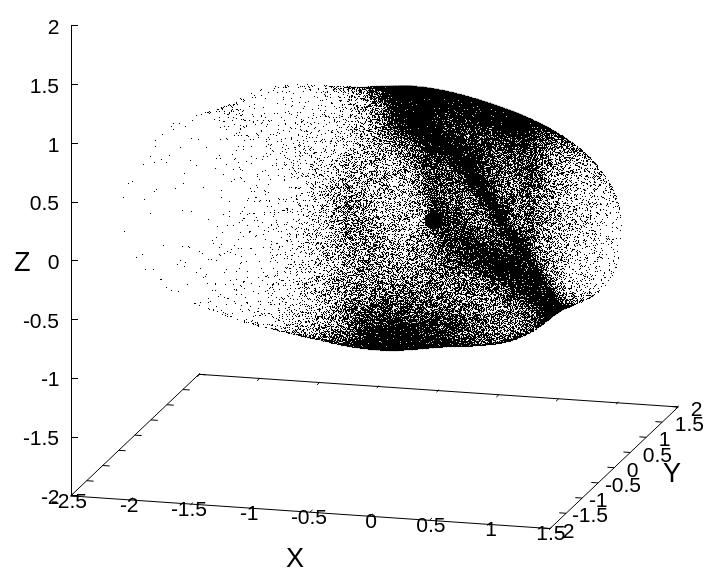}  
\includegraphics[width=8.9cm]{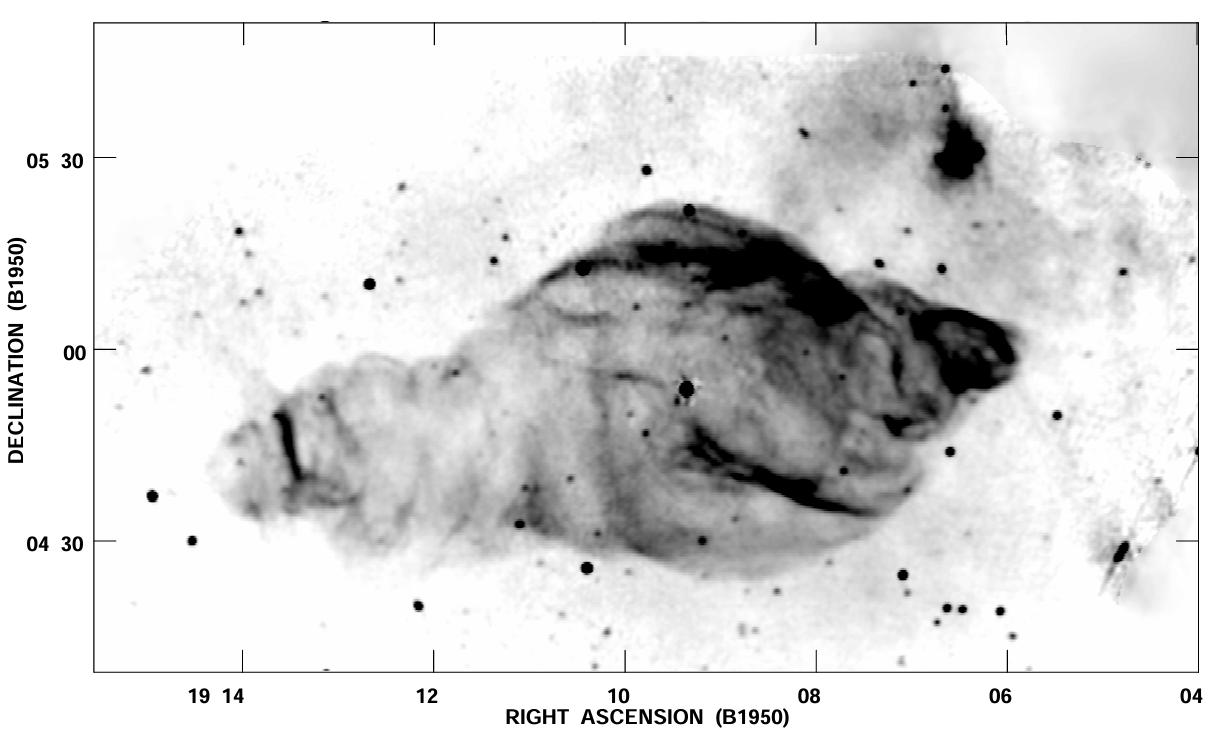}   
\end{center}
\caption{Explosion inside high-\va cigar-shaped cloud, mimicking the SNR, Manatee nebula (W50) in radio continuum.
The image was reproduced from  \citet{1998AJ....116.1842D}, https:// www.nrao.edu/pr/ 2013/w50/ .  
} 
\label{manatee}  
\end{figure*}

When the SN explosion takes place in a high-\va tube, the front expands in the direction of the axis of the tube, attaining a spindle shape, as shown in  figure \ref{manatee}.
Here, the magnetic tube is further superposed with oblique sheets of variable \va, which produce the oblique stripes and wavy deformation of the front.
The simulated shape mimics the Manatee nebula (W50), whose large-scale morphology  \citep{2017MNRAS.467.4777F} can thus be possibly explained by the interaction of the SNR exploded in a magnetic tube,
{
although more sophisticated model has been required to explain its extremely peculiar properties related to the central active source 
\citep{2021ApJ...910..149O,2023PASJ...75..338S}. 
}
The off-plane location and large angle of the major axis to the Galactic plane of the Manatee nebula \citep{1998AJ....116.1842D} suggests that the SNR exploded in a vertical magnetic tube inflating from the Galactic disc due to the Parker instability. 
The overlaid oblique stripes are also explained by corrugation of the front by sheer inhomogeneity of the medium, where the entire region is embedded in sheets obliquely tilted by $45\deg$ from the $x$ axis having sinusoidal variation of \va with $\delta \va\sim 0.01$ and $\lambda \sim 0.2$.

\ss{Corrugated filaments by sheet fluctuation}

\sss{Filamentary loop}

\begin{figure*}   
	\begin{center}  
(A) Filaments as corrugated surface: SNR 0509-67.5\\
\includegraphics[width=6.7cm]{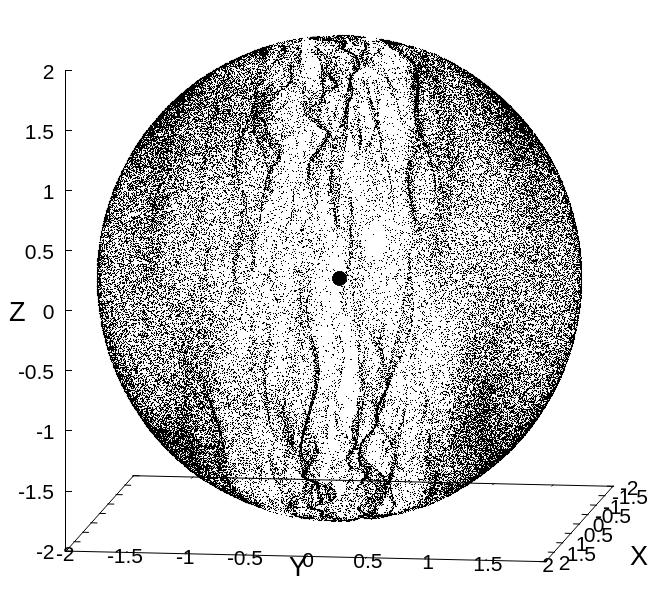}  
\includegraphics[width=7cm]{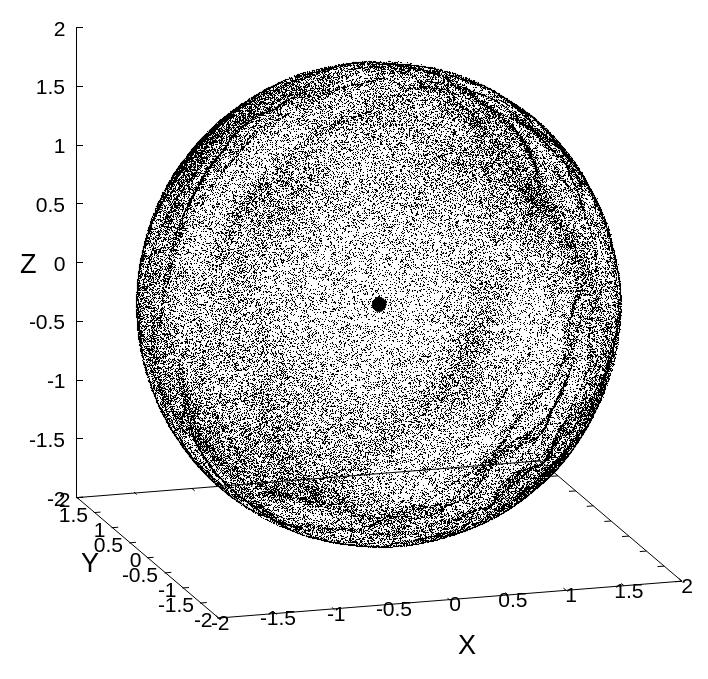}   
\\
(B) Filaments near the edge of Cygnus Loop\\
\includegraphics[width=7.cm]{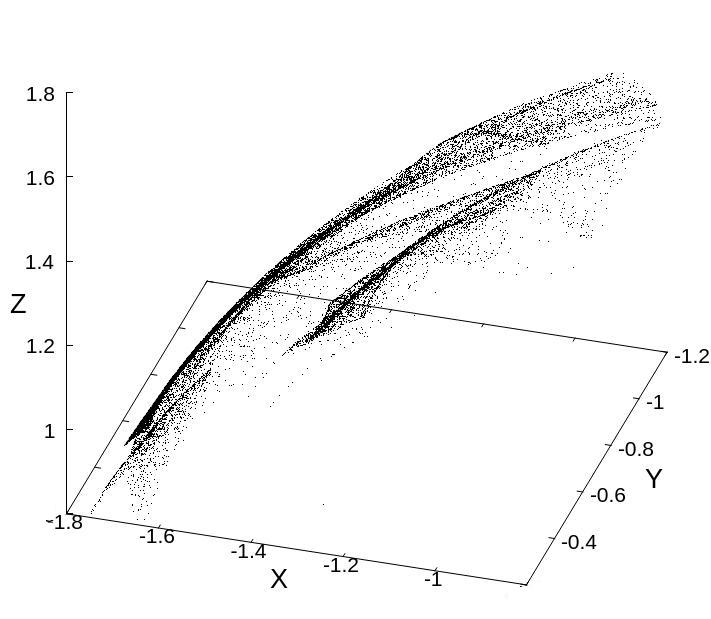}  
\includegraphics[width=7cm]{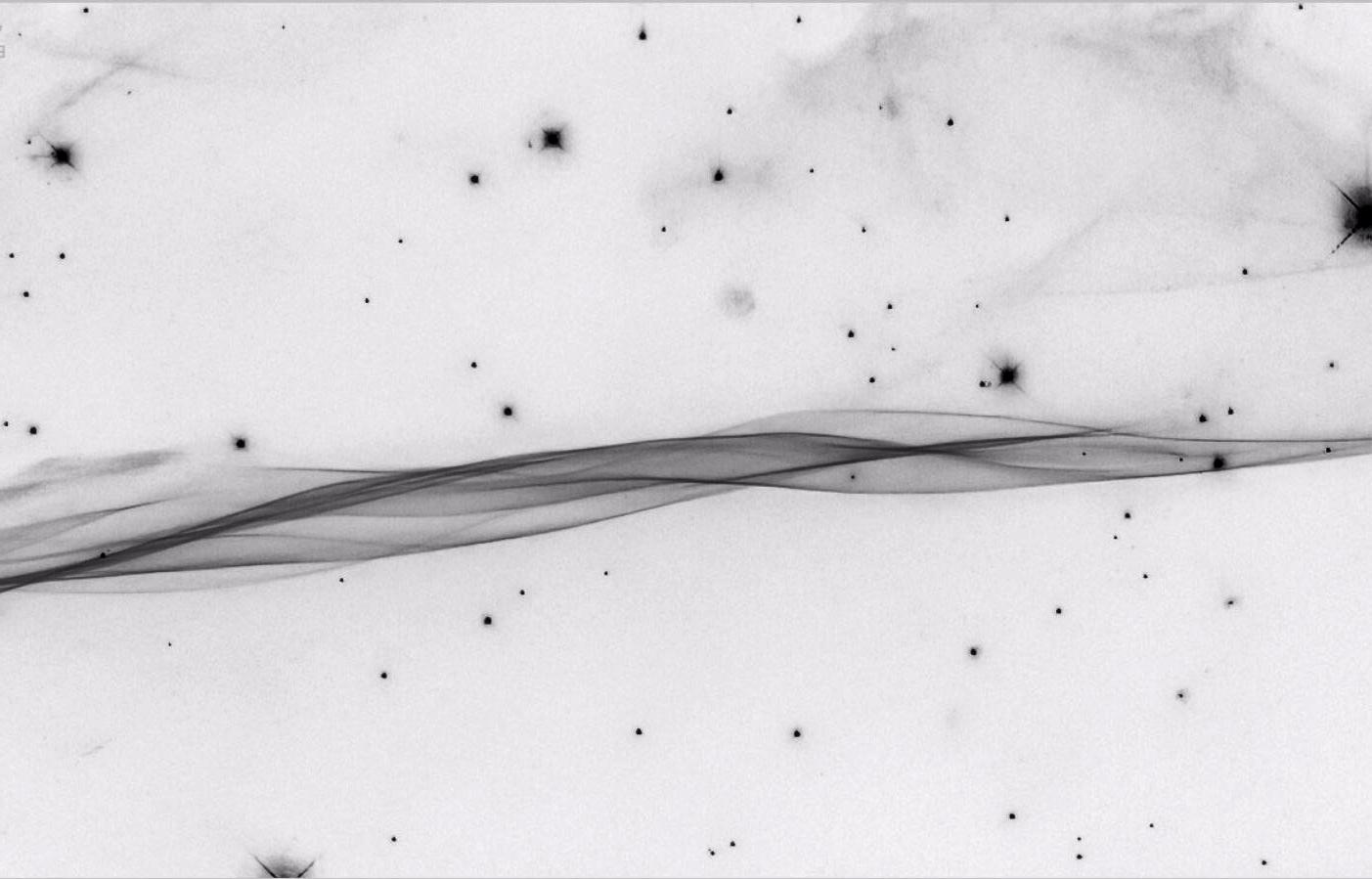}  
\end{center} 
\caption{{(A) Sheet filaments by tangential view of a corrugated surface near the edge of an almost perfect shell. 
(B) Close up of the filaments in Cygnus Loop \citep{2005AJ....129.2268B} (https://apod.nasa.gov/apod/ap200928.html 
) compared with the simulation.} }
\label{corrugation}   
\end{figure*}

If ISM has sheet-like fluctuations in one direction, the MHD front is disturbed to form wavy fluctuations near the equator parallel to the sheets' plane.
Figure \ref{corrugation} shows a case when the \va variation is sinusoidal in the $y$ direction by $\sim 1$ percent of the ambient value,
\be
V=V_0[1+\Sigma_j \epsilon_j \cos (2\pi y/\lambda_i)],
\ee
where $\epsilon_i \sim 0.01$ and $\lambda_i \sim 0.1$.

The perturbation influences the rays that propagate near the $(x,z)$ plane ($y\sim 0$) across the explosion center, yielding wavy front around the equator.
When the shell is seen from the pole direction ($y$ axis) axis, the projected shell exhibits wavy loop and arcs due to the tangential view of the corrugated surface of figure \ref{corrugation}) (A).
When the shell is viewed from the $x$ direction, filament structure is observed as equatorial filaments across the center.

\sss{Thin filaments in Cygnus Loop}

Filamentary structure in the tangential projection of the outer shell is more ubiquitous in older SNRs.
The very thin filaments in the Cygnus Loop as shown in the right panel of figure \ref{corrugation} (B) \citep{2005AJ....129.2268B} can be naturally reproduced by a tangential view of corrugated front, as simulated by the MHD-wave model shown in figure \ref{corrugation} (B).

\section{Turbulent and peculiar morphology}
\label{sec4}

\ss{Randomly distributed high- and low-\va clouds}

When the MHD wave front expands into a turbulent medium, the front is deformed to produce clumps, filaments, arcs, loops and holes.
We simulate three cases with randomly distributed 100 Gaussian clouds having  random sizes, amplitudes, and positions in the plotted frame. 

\begin{figure*} 
\begin{center}      
(A) High-\va random clouds, $t=1$ and 2 \hskip 3cm B. Low-\va clouds\\
\includegraphics[width=7cm]{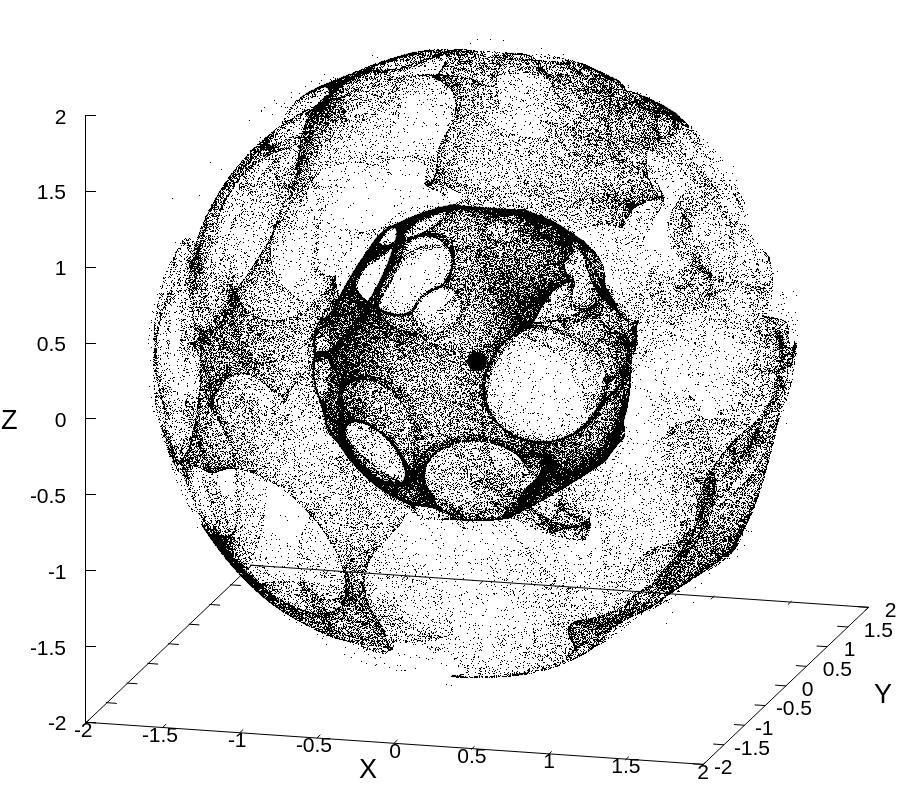} 
\includegraphics[width=7cm]{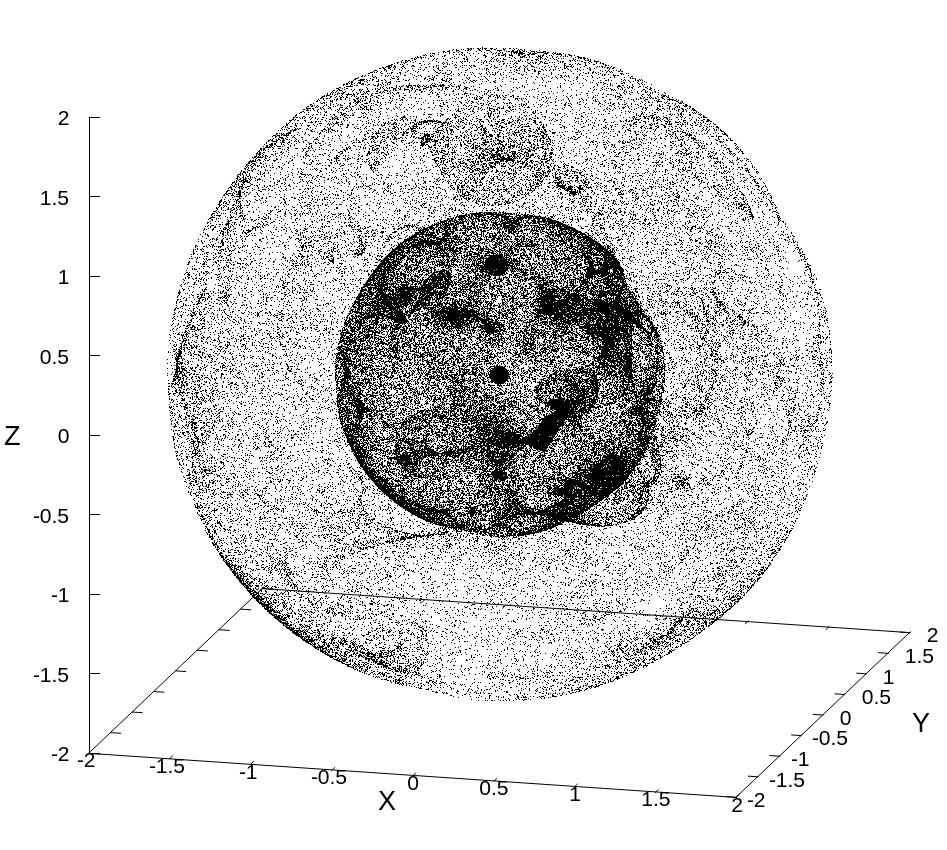}  \\
(C) Low-(blue) and high-(red) \va clouds \hskip 2cm D. MHD front  \\ 
\includegraphics[width=7cm]{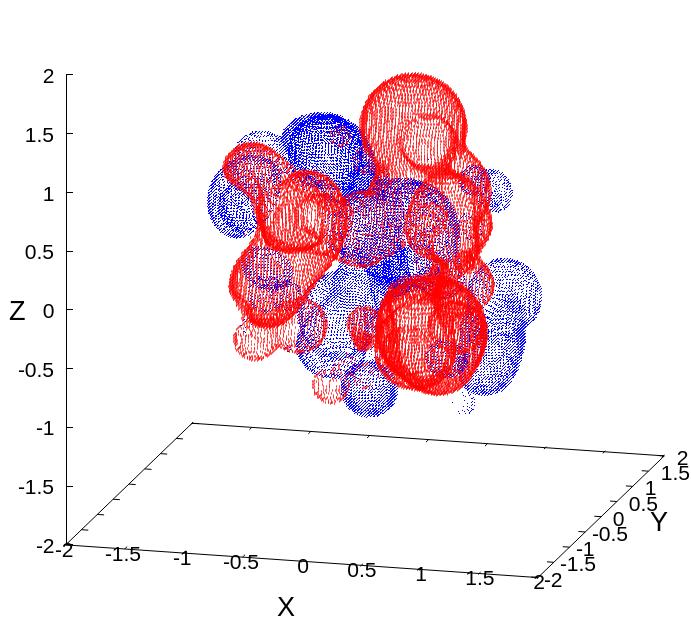} 
\includegraphics[width=7cm]{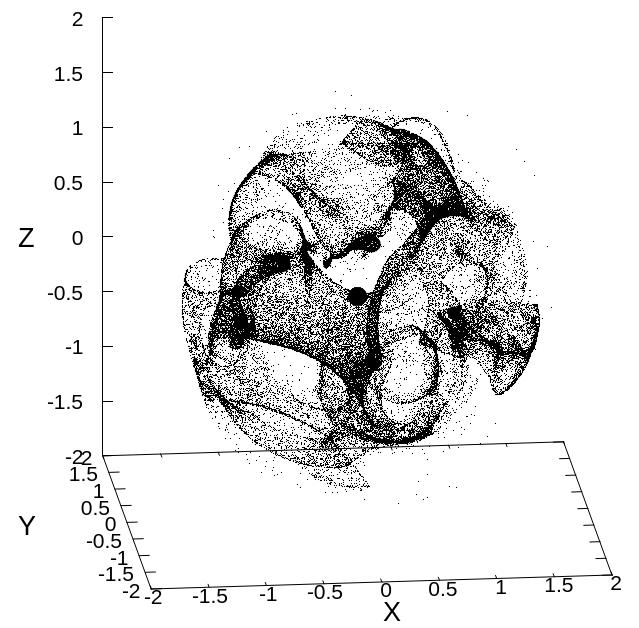}   
\end{center}
\caption{(A) MHD front in random (turbulent) Gaussian clouds:
(B) Same, but in low-\va clouds. 
(C) Randomly distributed 100 low- and high-\va clouds. Blue clouds have low-\va, and red high-\va. Note, enlarged scale in this panel.
(D) MHD front in the medium given by panel C.} 
\label{random} 
\end{figure*}

Figure \ref{random} compares the time evolution of MHD fronts at $t=1$ and 2$t_{\rm unit}$, expanding into turbulent medium filled with randomly distributed high-V (left) and low-V (right) Gaussian clouds.
High-V clouds deform the shell more strongly than low-V clouds.

If the medium is filled with high-\va clouds, the front is largely deformed to exhibit filaments, arcs, and holes as shown in figure \ref{random} (A).
When the interstellar medium is composed of low-\va random clouds, 
the front is deformed to make randomly focusing, concave, and locally amplified clumps, as shown in  figure \ref{random} (B).
{If the medium is mixed with low- and high-V clouds, the front results in a complicated deformed shell with the mixture of concave clumps, holes, and filaments as shown in figure \ref{random} (D).    }
Here, the \va distribution in the ISM is given superposition of random Gaussian \va clouds expressed by
\be V=1+\Sigma_i \delta V_i \ee
\label{eqran}
with
\be
\delta V_i=v_i \exp{\left(-\frac{(x-x_i)^2+(y-y_i)-2+(z-z_i)^2}{w_i^2}\right)},
\ee
where 
\be      w_i=0.05+0.1\times \mathcal{R}(4,i), \ee
\be      x_i=2(\mathcal{R}(1,i)-0.5),\ee
\be      y_i=2(\mathcal{R}(2,i)-0.5),\ee
\be      z_i=2(\mathcal{R}(3,i)-0.5),\ee 
\be      v_i=2(\mathcal{R}(5,i)-0.5),\ee 
and $\mathcal{R}(j,i)$ represents random number between 0 and 1, as shown in figure \ref{random} (C).

\begin{figure*} 
\begin{center}      
(A) Filaments and loops by high- and low-\va clouds
(B) The Vela nebula\\
\includegraphics[width=8cm]{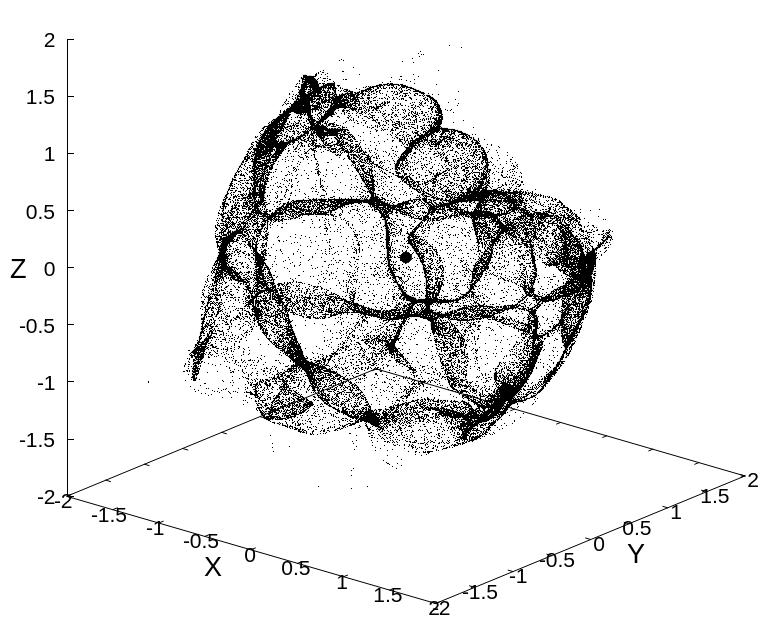}  
\includegraphics[width=6.5cm]{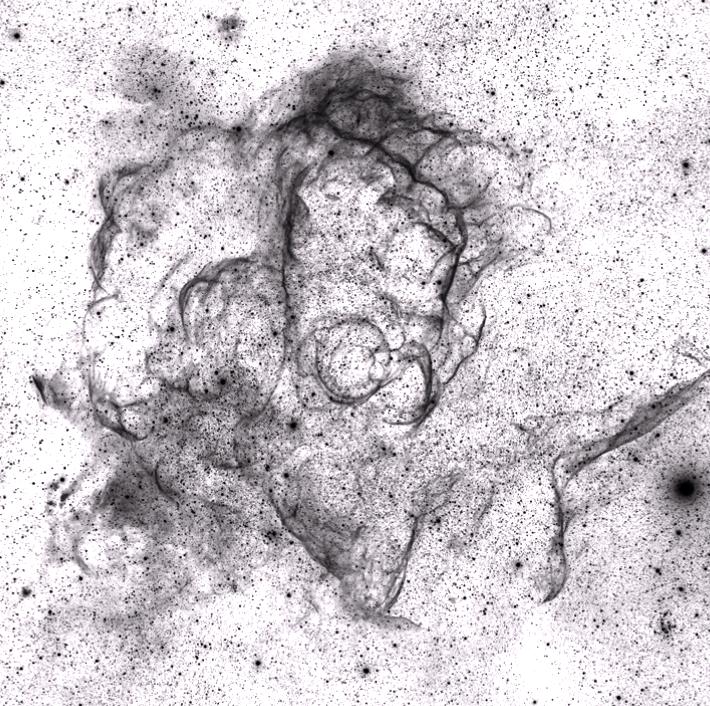}  
\end{center}
\caption{(A) MHD-wave front expanding in randomly distributed high- and low-\va clouds.
(B) The Vela nebula reproduced from
https://apod.nasa.gov/apod/ap100910.html. 
}
\label{tycho}
\label{vela} 
\end{figure*}

Expansion into a turbulent ISM filled with high-V clumps results in a front filled with cell-like clumps of arc-shaped filaments.  
As the wave front expands further, the turbulent ISM with high-V clumps causes more disturbed structure.
The Vela nebula would be an example of such deformed front, where a number of filamentary arcs and loops are observed as shown in figure \ref{vela} (B).

\ss{Expansion into sheets and tubes: oblique stripes} 

\begin{figure*} 
\begin{center}       
(A) Sinusoidal sheets ~~~~~~~
(B) Sinusoidal filaments\\
\includegraphics[width=7cm]{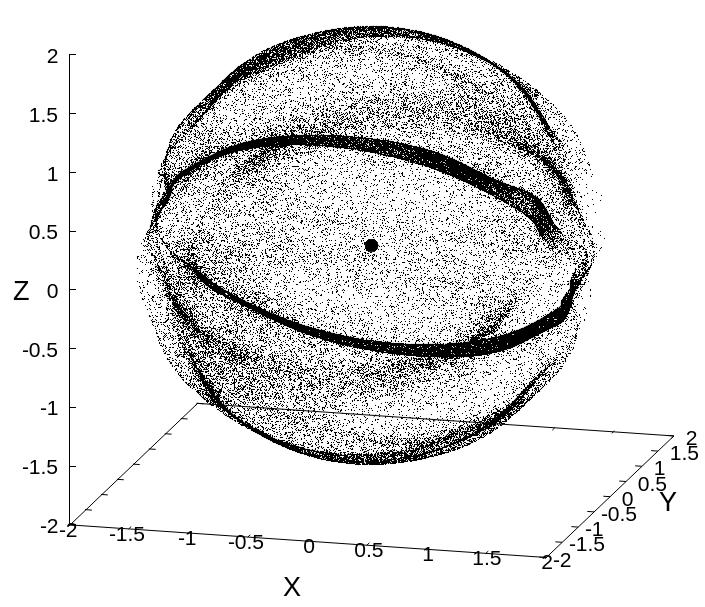}  
\includegraphics[width=7cm]{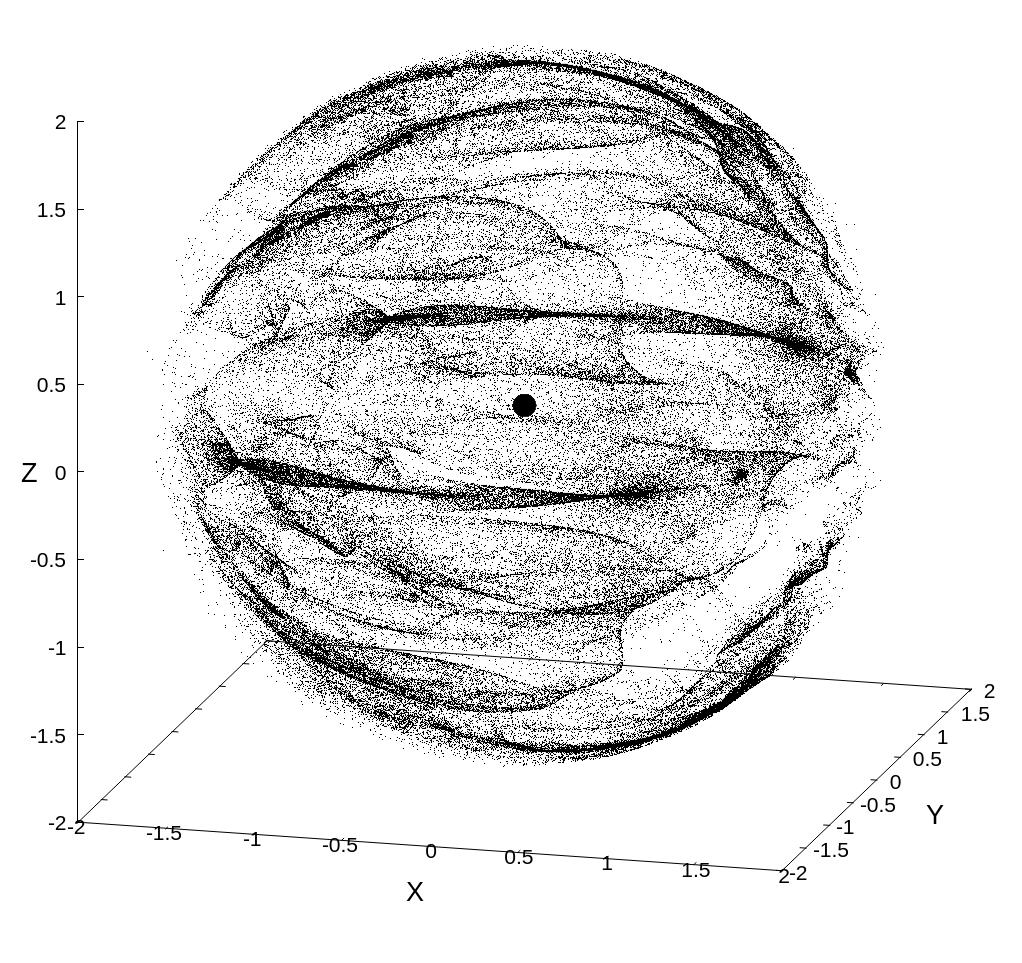}   
\end{center}
\caption{{MHD fronts interacting with sinusoidal \va sheets and tubes. 
} }
\label{ctb1}   
\end{figure*}

Plane parallel sheets with sinusoidal \va variation creates longitudinal stripes.  
If it is superposed by sinusoidal variation in the $y$ and $z$ directions, mimicking high- and low-\va tubes, the longitudinal stripes appear to cover the surface as shown in figure \ref{ctb1}, where the wavelengths are taken as $\lambda_y=\lambda_z=1$.   
 
\begin{figure*} 
\begin{center}         
(A) High-\va filaments vs Low-\va filaments\\
\includegraphics[width=6.7cm]{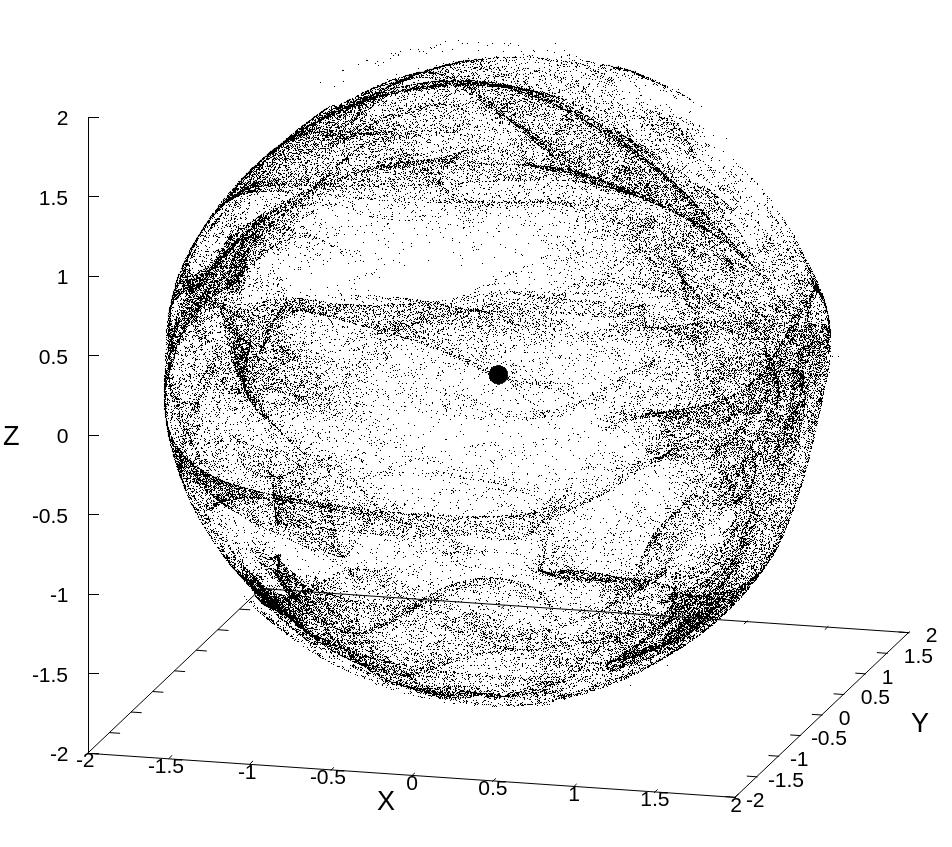}   
\includegraphics[width=6.7cm]{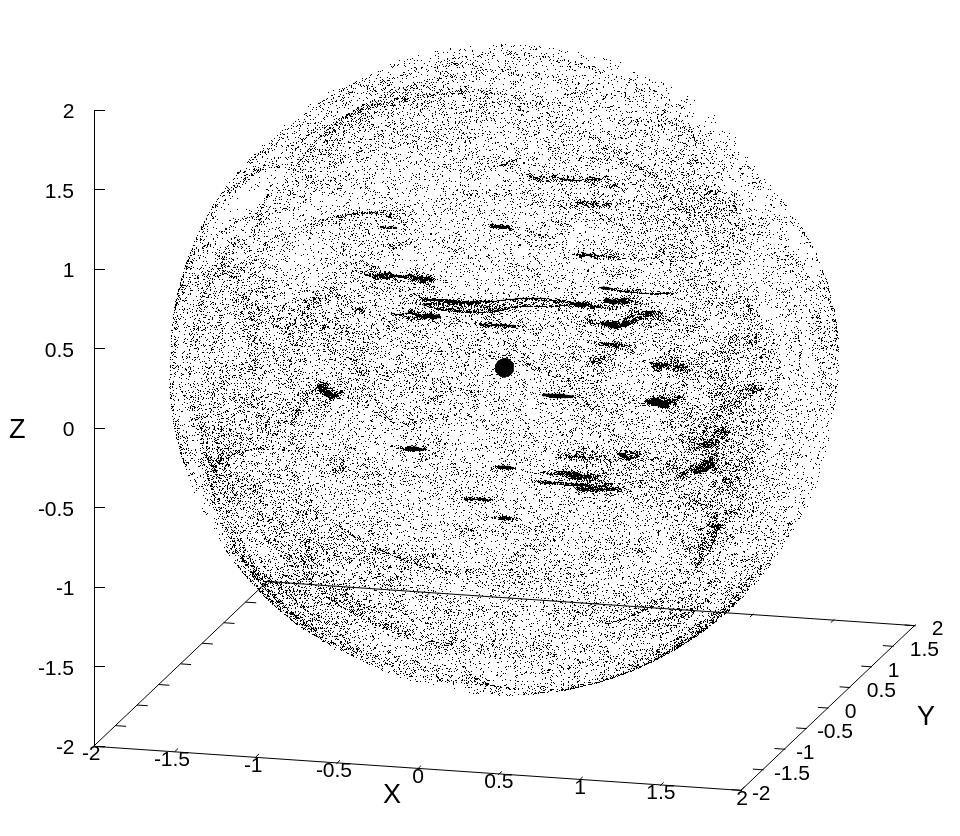}   \\ 
(B) Side view vs  Pole view \\
\includegraphics[width=6.7cm]{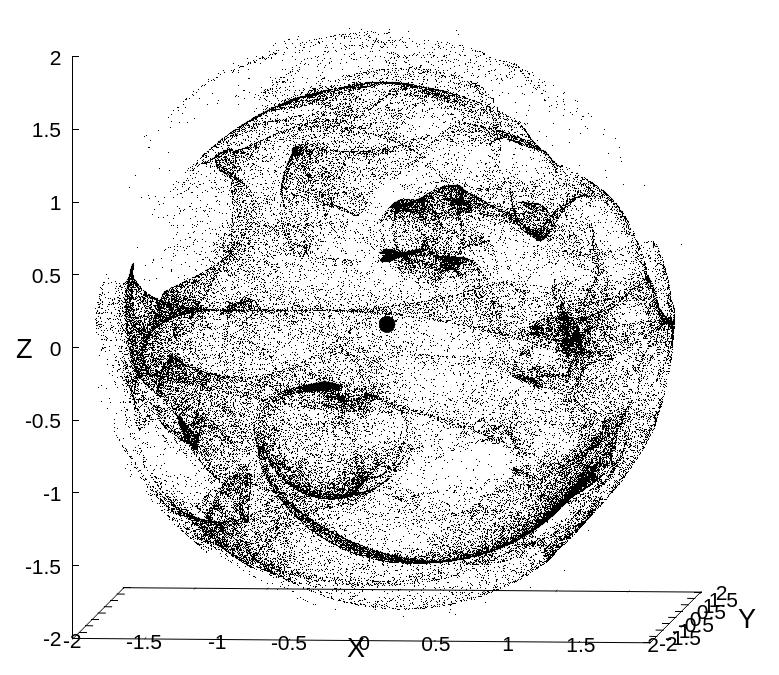}  
\includegraphics[width=6.7cm]{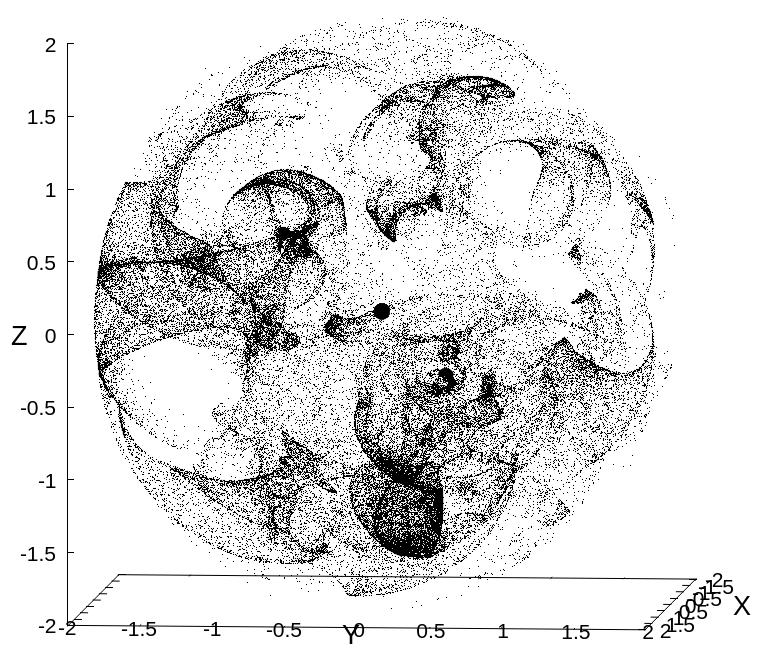}  
\end{center}
\caption{(A) Formation of filamentary {MHD-wve fronts} by encounter with elongated low- and high-\va strings of clouds. 
(B) Encounter with elongated clouds, but shows the difference of the appearances depending on the viewing angle. The left panel shows a perpendicular view, while the right is an end-on view parallel to the filaments. 
}
\label{filaments}  
\end{figure*}

{MHD wave front} expanding into ordered ISM with directional filaments with high-\va results in a shell covered by filaments and stripes parallel to the perturbing ISM. 
Low-\va filaments result in less deformed filamentary surface.
Figure \ref{filaments} (A) shows the simulated results.

However, if the deformed filaments are viewed from the direction parallel to the filament axes, the deformed front resembles the result for random high-\va clouds.
Figure \ref{filaments} (B) shows the same result for filamentary ISM turbulence viewed from two different directions: 
The left panel  is a view perpendicular to the filaments, and the right is parallel.
The projection parallel to the filaments resembles to random-cloud cases.

\ss{Morphological evolution}
 
\begin{figure} 
\begin{center}      
\includegraphics[height=5.5cm]{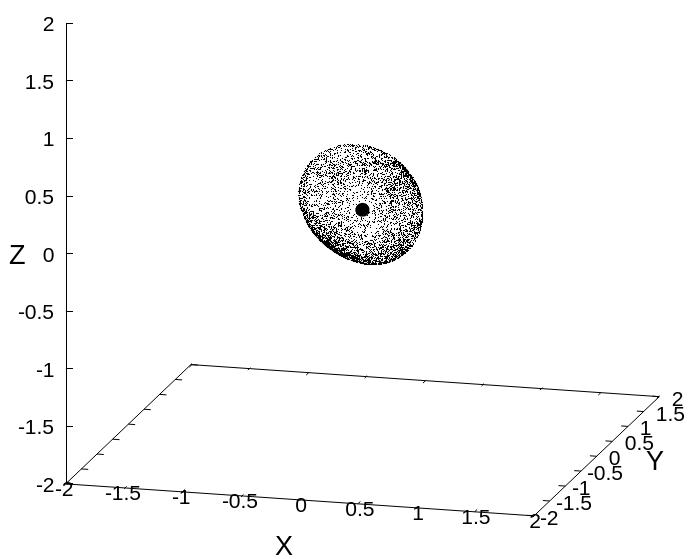} 
\includegraphics[height=5.5cm]{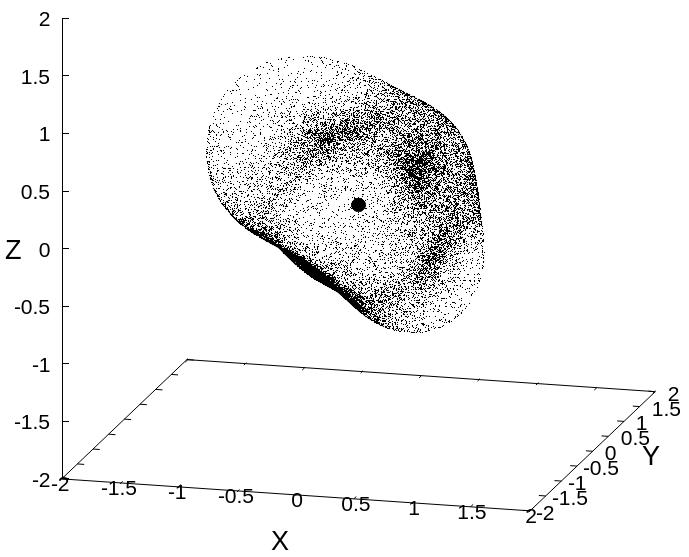}   
\includegraphics[height=5.5cm]{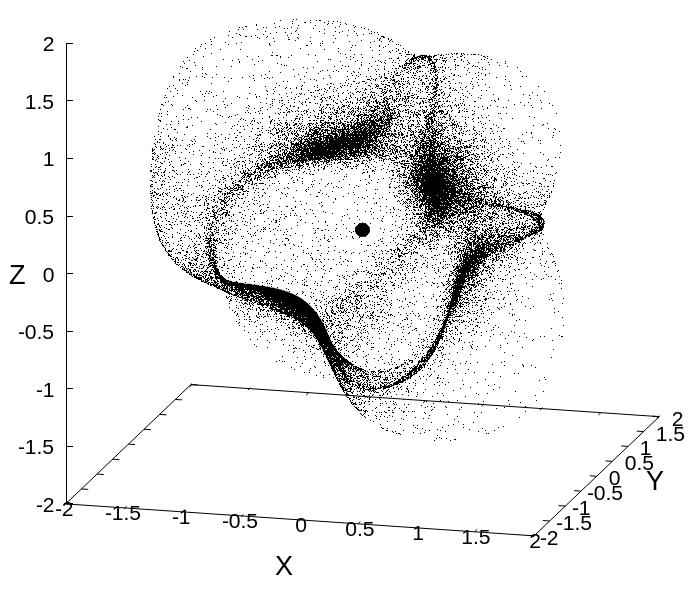} 
\includegraphics[height=5.5cm]{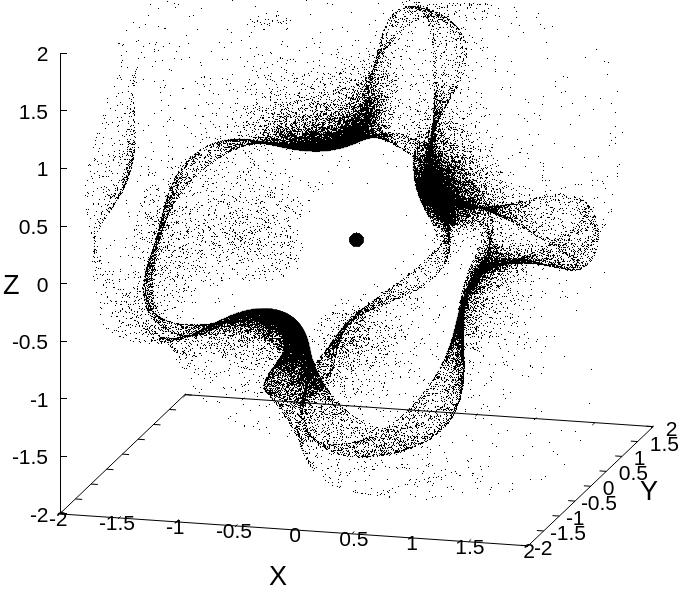}    
\end{center}
\caption{Morphological evolution of {MHD-wave front} expanding into a  medium with sinusoidal \va variation, plotted at $t=0.5$, 1, 1.5 and 2.} 
\label{evolution1} 
\end{figure}

\begin{figure}
\begin{center} 
\includegraphics[height=5.5cm]{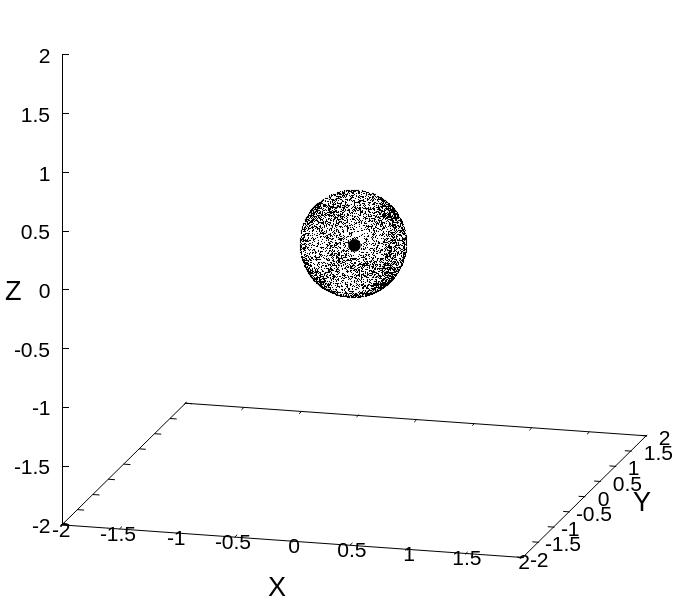} 
\includegraphics[height=5.5cm]{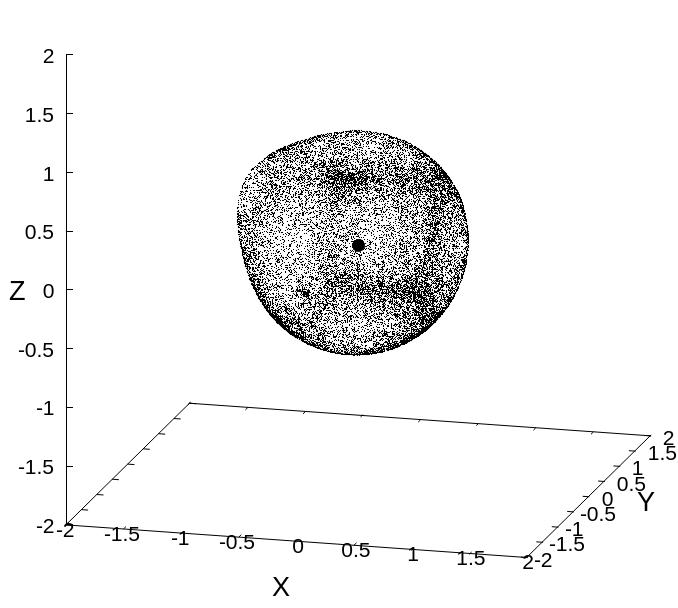}   
\includegraphics[height=5.5cm]{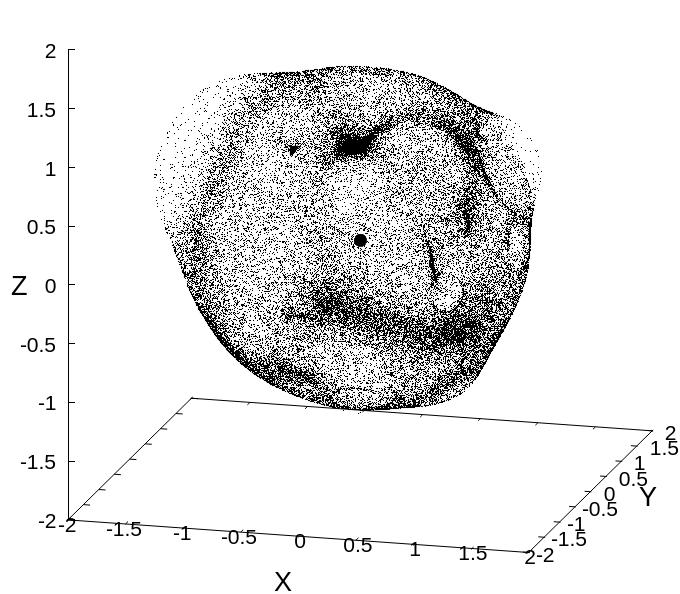} 
\includegraphics[height=5.5cm]{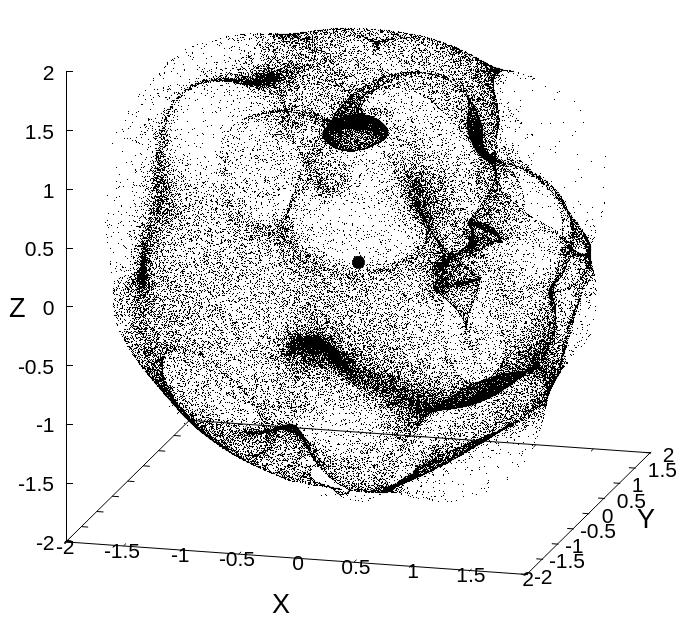}    
\end{center}
\caption{Morphological evolution of {an expanding MHD-wave front} into a turbulent medium filled with randomly distributed high- and low-\va clouds plotted at $t=0.5$, 1, 1.5 and 2.} 
\label{evolution2} 
\end{figure}

We have so far presented snap shots of the simulated result.
Besides snap shots, it is also interesting to see how the morphology evolves during the expansion. 
In figure \ref{evolution1} and \ref{evolution2} we show the time variation of an expanding MHD wave front in a turbulent medium filled with sinusoidal and randomly distributed high- and low-\va clouds. 
 
The shell expands spherically in the initial stage, until the size becomes comparable to the scale size of the turbulence.
After a while, the front is deformed to exhibit asymmetric and elongated shell due to the spatial gradients in the \va distribution.
As it further expands, the deformation is amplified to attain a turbulent morphology covered by randomly distributed filaments, arcs, holes, and concave focal clumps.  

\section{{Qualitative understanding of peculiar morphology of SNRs}}
\label{sec5}

\ss{Cygnus Loop: Encounter with low-\va bars} 

The Cygnus Loop is known as the most typical shell-type SNR, but it also shows peculiarly open and complicated morphology divided by a bunch of long straight filaments crossing the center \citep{2018MNRAS.481.1786F}.
It consists of largely deformed arcs and filaments:
The Veil Nebulae (NGC 6992) composes the eastern limb of a bunch of bright arcs and filaments, and the western limb also composes a bunch of veil nebulae and extends to the south.
The entire SNR is largely deformed and is opened toward the south, making a long protrusion.
The most spectacular structure is the straightly extending bunch of vertical (north-south) filaments, which separate the whole shell into eastern and western halves.

\begin{figure*} 
\begin{center}       
{(A) Stereograph of {MHD-wave front, qualitatively mimicking the} Cygnus Loop}\\
\hskip -1.5cm
{{\bf Left eye}} \hskip 4cm {\bf Right eye}\\
\includegraphics[width=13cm]{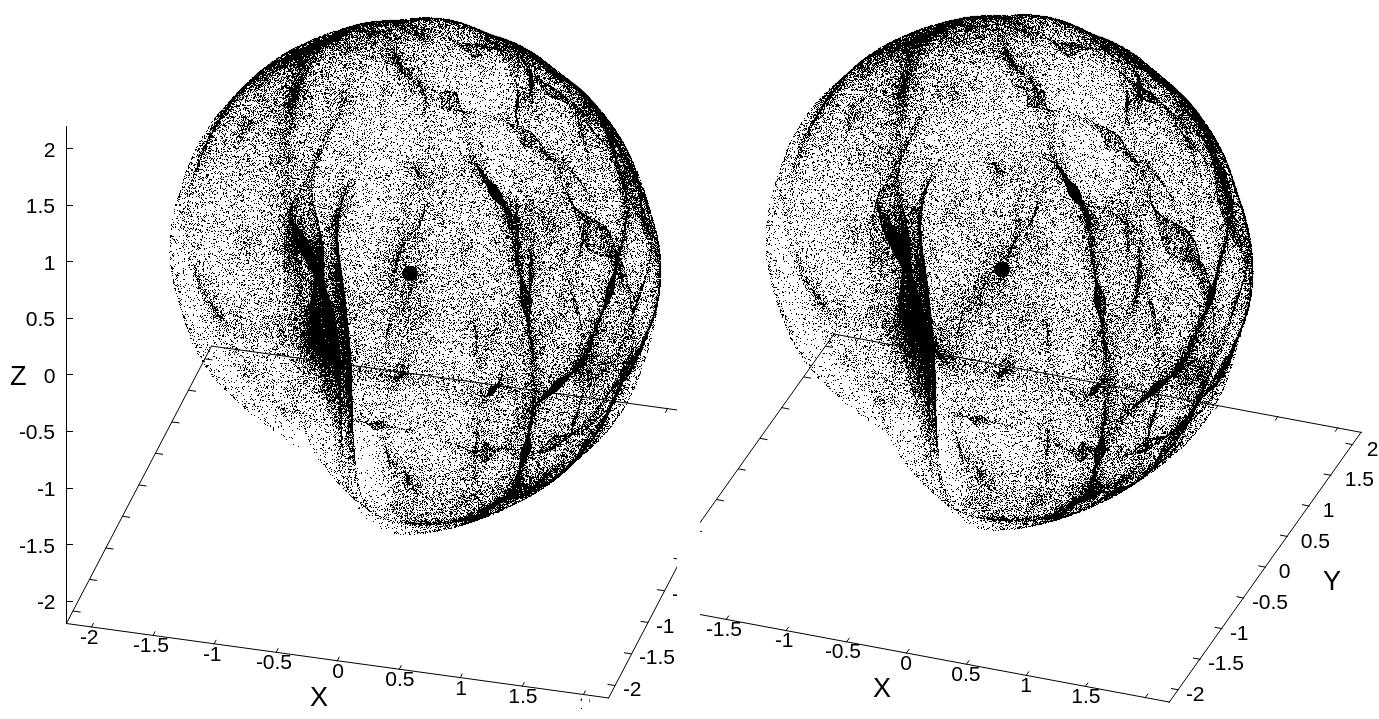}  
\includegraphics[width=2.5cm]{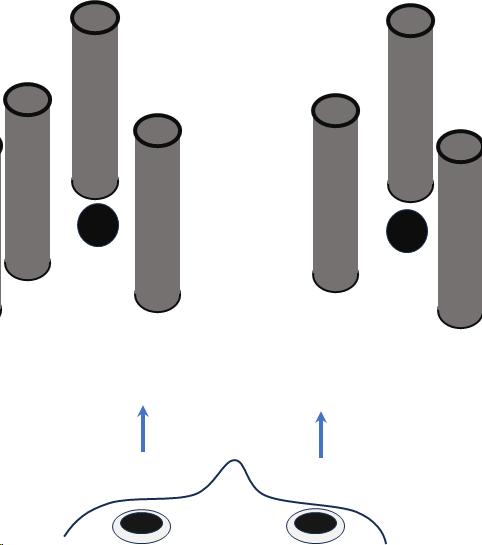}  \\
\vskip 3mm
(B) Encounter with low- and high-\va clouds \hskip 2cm (C) View from the top.\\
\includegraphics[width=7cm]{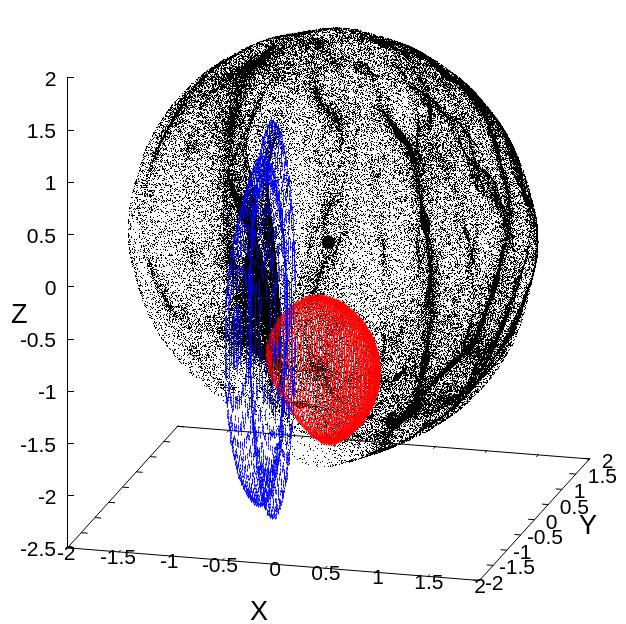} 
\includegraphics[width=6.5cm]{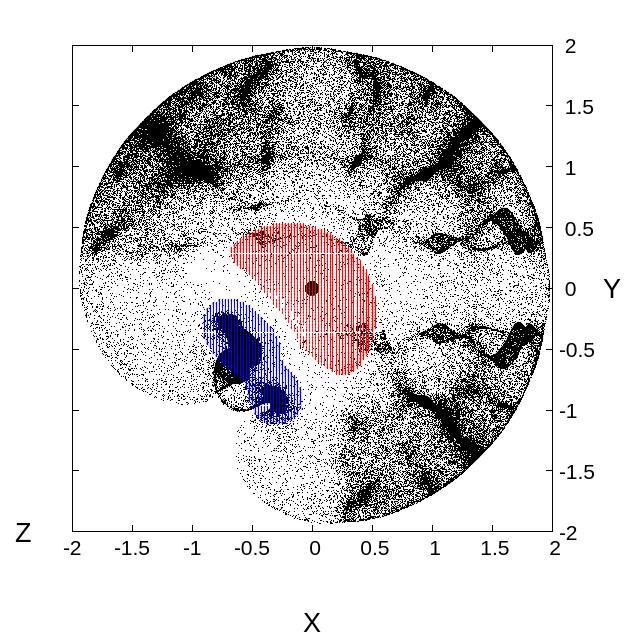}\\
(D) Cygnus Loop\\
\includegraphics[width=7.5cm]{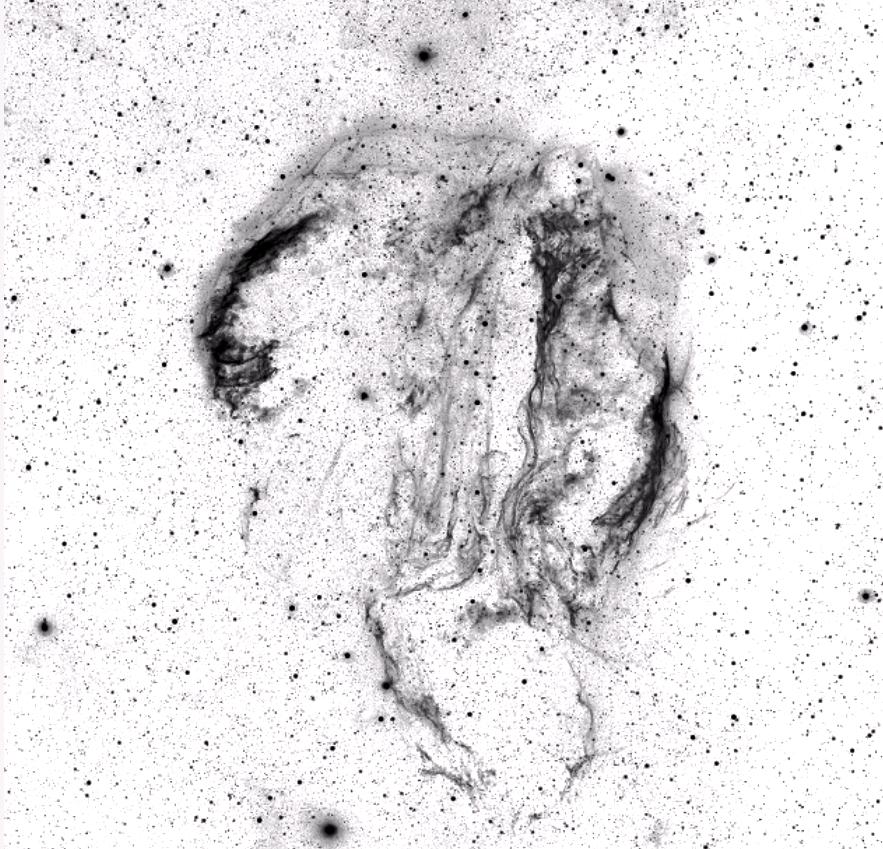}   
\end{center}
\caption{(A) Stereographic presentation of the simulated MHD front for the Cygnus Loop. {See the left and right panels individually with the left and right eyes, respectively.}
(B) Encounter with vertical bar-shaped clouds of low \va, resulting in a long, extended bunch of filaments, and with a high-\va cloud in the southern edge, mimicking the SNR, Cygnus loop.
(C) Same, but a view from the top (north).
(D) Optical image of the Cygnus Loop 
(North to the top:  https://apod.nasa.gov/apod/ap191031.html ).
} 
\label{cygnus} 
\end{figure*}

{These striking features have been well simulated by a lopsided outburst of the progenitor star in a uniform ISM
\citep{2017MNRAS.464..940F}.
We here try to simulate such peculiar morphology qualitatively by the lensing effect, where an initially spherical shell expands into an inhomogeneous medium.
}
We assume that the front encounters a vertical (NS-oriented) long bar-shaped cloud with low-\va and a high-\va cloud near the southern edge.
Figure \ref{cygnus} shows the results compared with a photograph of the SNR, where blue and red dots represent the low and high \va clouds, respectively.
In the top panel we show a stereoscopic view of the simulated results, one panel rotated $10\deg$ from the other perpendicular to the line of sight to explore the 3D structure of the SNR.

\ss{The $\tau$ nebula: two orthogonal low-\va bars}

\begin{figure*} 
\begin{center}         
(A) {MHD wave front } ~~~~~~~~~ 
(B) Two orthogonal bar  clouds ~~~~~~
(C) The $\tau$ nebula.\\
\includegraphics[width=6cm]{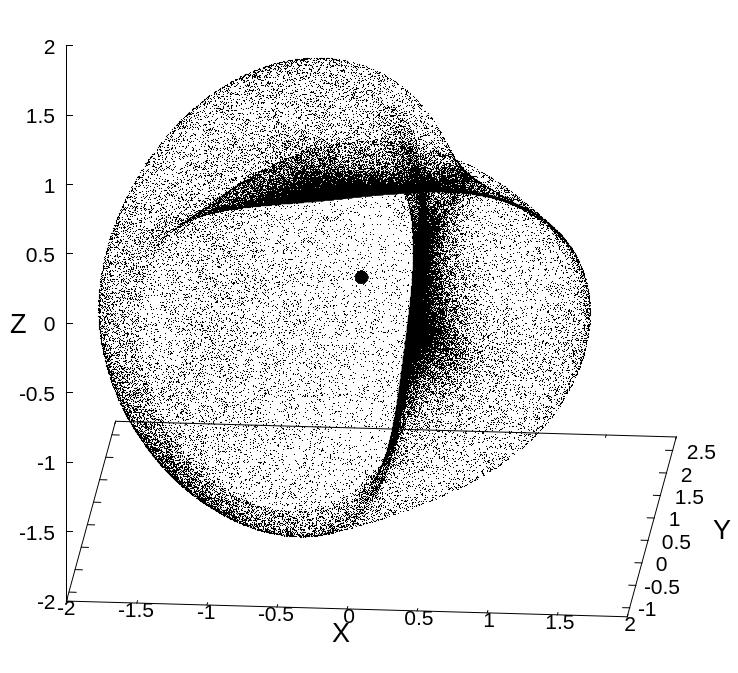}  
\includegraphics[width=6cm]{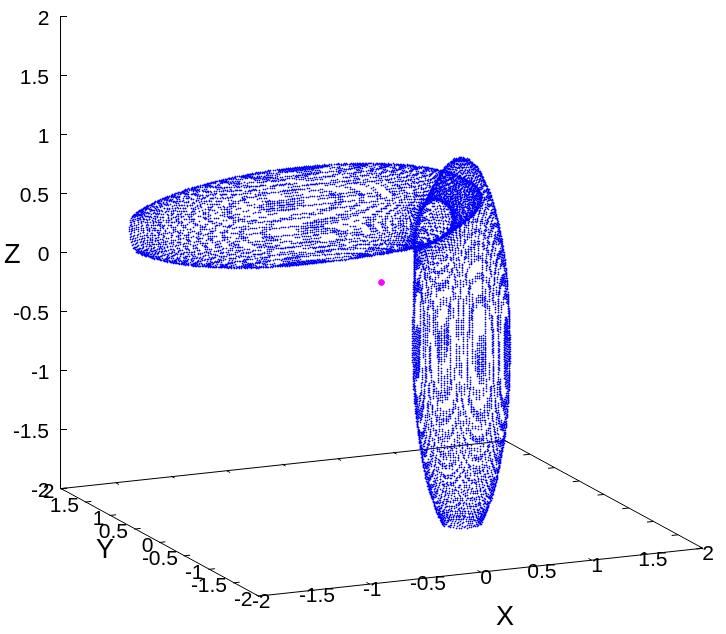}   
\includegraphics[height=4.5cm]{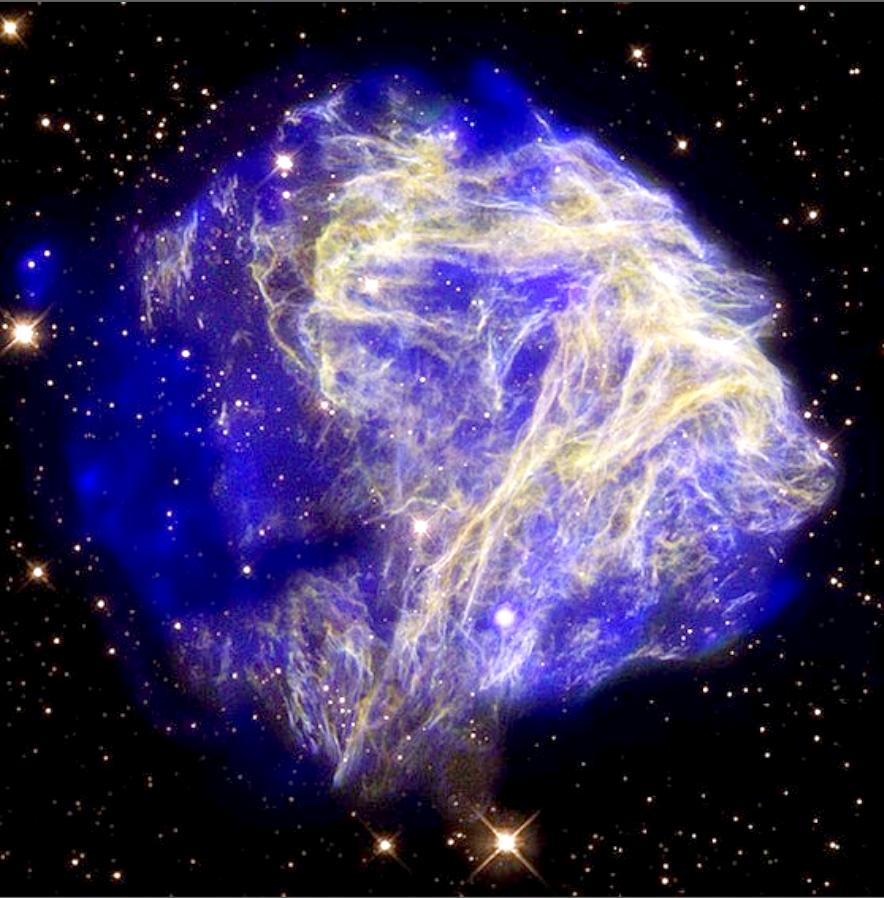}  
\end{center}
\caption{{Deformed MHD-wave front (A) by an encounter with (B) two orthogonal bar-shaped clouds having low \va, qualitatively mimicking (C) the $\tau$ nebula, N49 in LMC} 
( https://hubblesite.org/contents/media/images/  
).
} 
\label{n49} 
\end{figure*}

The $\tau$ nebula (N49/DEM-L-90) in the LMC is an SNR with the most strange morphology, exhibiting two orthogonal bunches of straight filaments.
In our simulation, such morphology is understood as a result of an encounter of the MHD front with two orthogonal bar-shaped clouds with low-\va.
Figure \ref{n49} shows the simulation result and assumed low-\va clouds, as compared with the HST optical image.

\ss{SNR S147: loops, arcs, and converging filaments} 

\begin{figure*} 
\begin{center}         
(A) Cloud distribution around S147 \hskip 2cm (B) Approximate result.\\
\includegraphics[width=5.5cm]{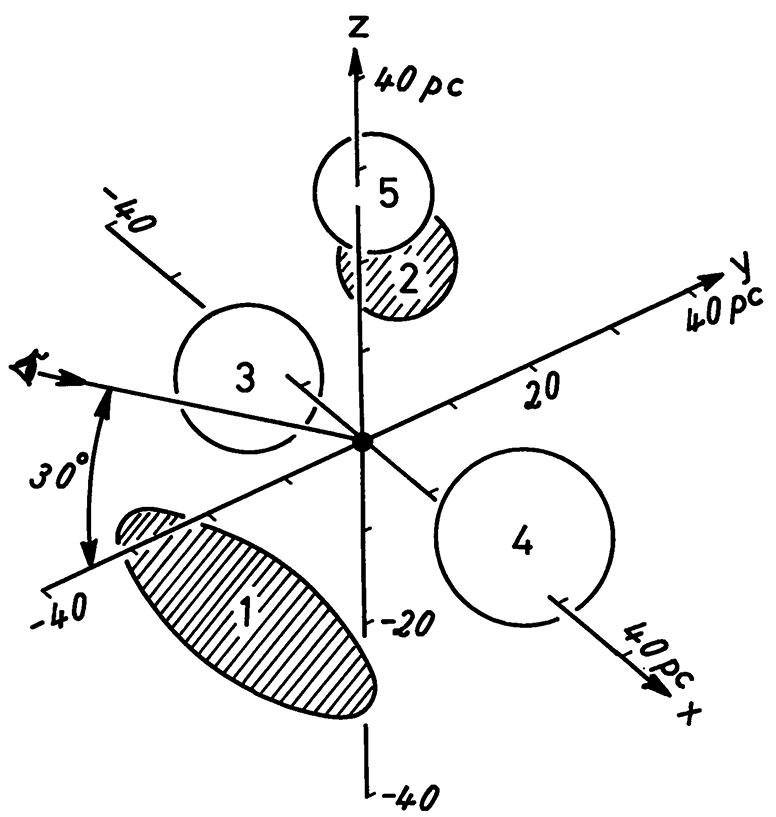}  \hskip 5mm
\includegraphics[width=6cm]{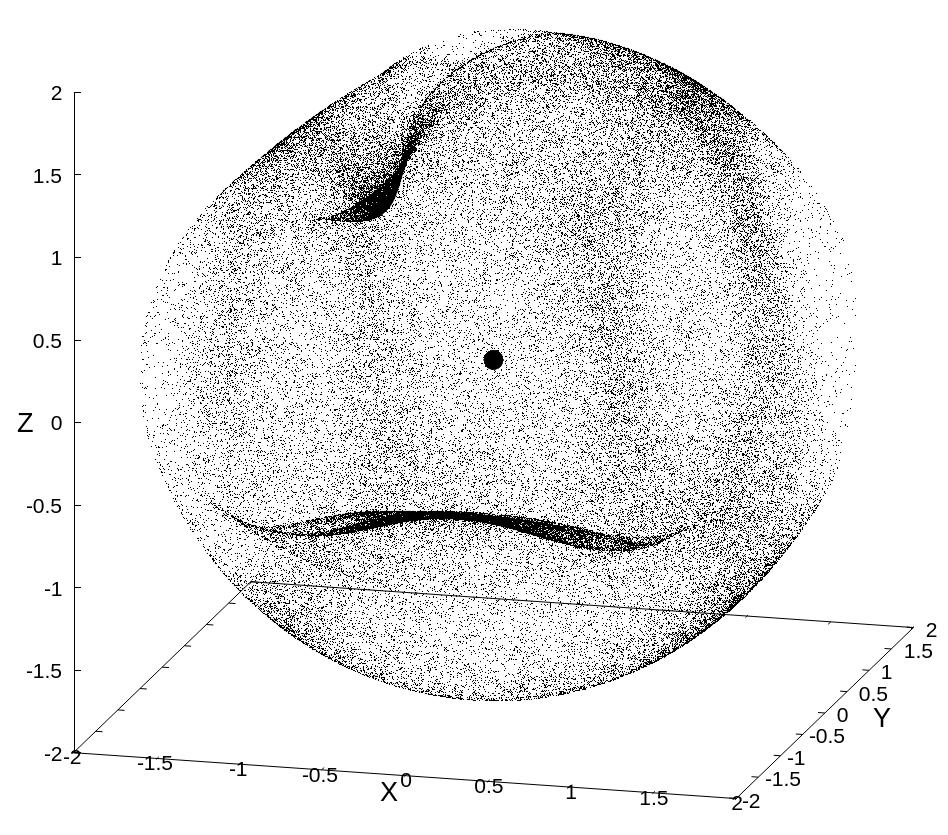}    \\
(C) MHD simulation of S147 \hskip 5cm (D) Same, but in fits image\\
\includegraphics[height=7.5cm]{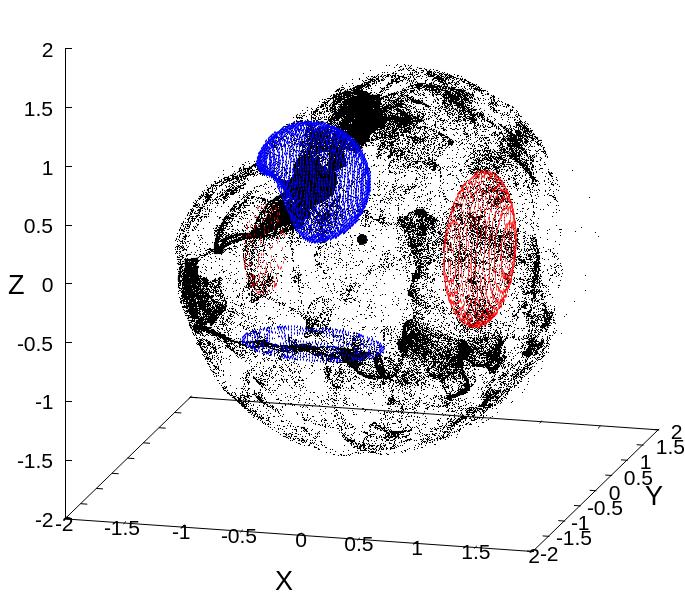}      
\includegraphics[height=7cm]{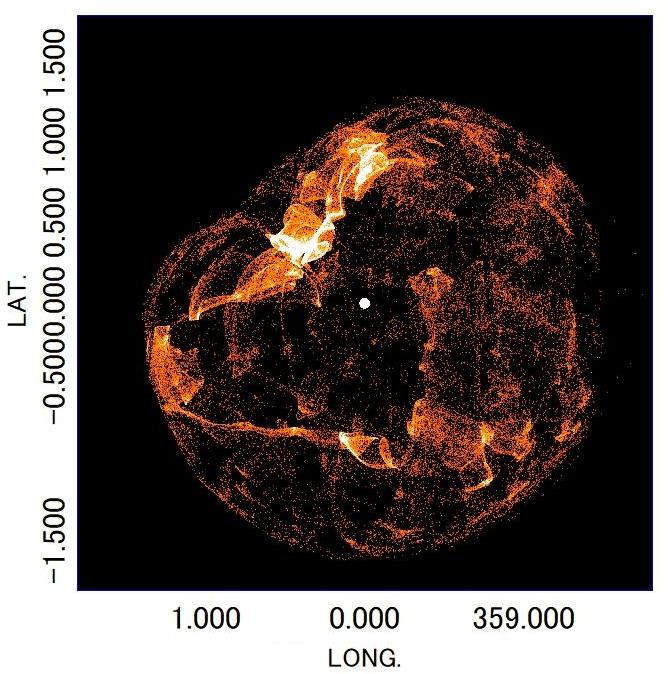}     \\
(E) S147 (North to top-left)\\  
\includegraphics[height=6cm]{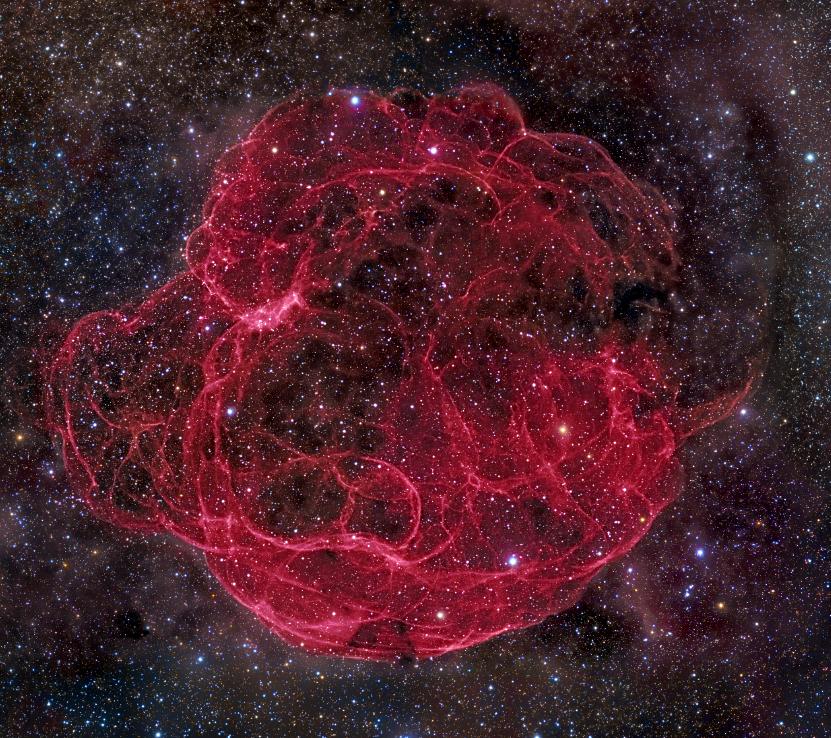}    
\end{center}
\caption{(A) Approximate cloud distribution and (B) an early simulation result for S147 \citep{1978A&A....67..409S}.
(C) Filamentary loops, arcs, and concave focusing clumps as the result of encounter with low- and high-\va clouds as marked by the bule and red dots, mimicking 
the SNR S147. 
(D) shows the same simulation, but by a fits-formated image. 
(E) Photo of S147 reproduced from https://apod.nasa.gov/apod/ap121009.html
}  
\label{s147}  
\end{figure*}

The old shell-type SNR S147 has been simulated in our early paper \citep{1978A&A....67..409S} based on the optical image \citep{1973ApJS...26...19V}.
In figure \ref{s147} we reproduce our early work by the present simulation.
We further performed higher-resolution simulation, and show the result in the lower panels of the figure in comparison with an optical image of S147 in the H$\alpha$ line.

Low-\va clouds 1 and 2 causes the concave filaments due to the focusing effect. 
High-\va clouds 3 and 4 leads to the convex deformation in the eastern and western edges.

\begin{figure*} 
\begin{center}       
(A) Simulated NE clump of S147 \hskip 2cm (B) NE clump of S147\\ 
\includegraphics[height=8cm]{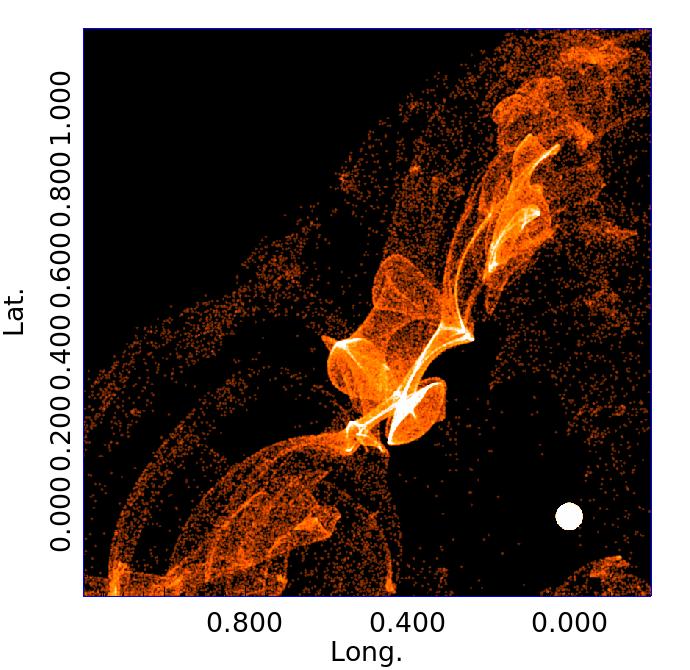}
\includegraphics[height=7.7cm]{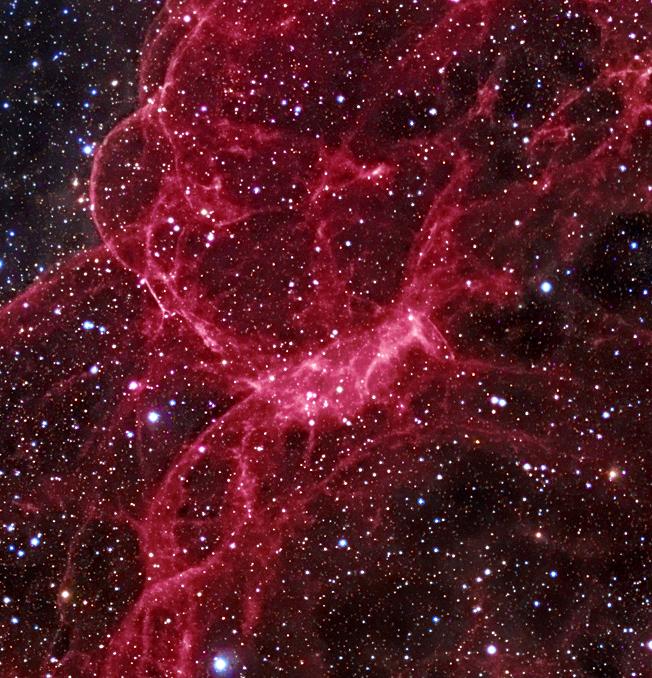}  
\end{center}
\caption{(A) Same as figure \ref{s147} for S147, but enlarged for the NE complex, which is a focusing clump by a big low-\va cloud, comapred with (B) the enlarged image of S147 (https://apod.nasa.gov/apod/ap121009.html)
}  
\label{s147up}  
\end{figure*}

Figure \ref{s147up} enlarges the NE filamentary region compared with the simulation.
Due to the focusing effect by the low-\va cloud, wave front is strongly deformed and many converging waves produce a condensation of overlapped filaments, leading to a concave bright clump.

\section{Discussion}
\label{secdiscussion}

\ss{{Limitation of the method and speculative application of the model to SNRs}}

By definition, the present MHD wave simulation cannot be applied to the early phase of SNR expansion, when the mass of plowed gas is less than the mass of the progenitor. 
Also, the method is not applicable to SNRs with expansion velocity much faster than the \alf velocity with $V_{\rm expa}>> 10$ \kms as well.
However, we argue that the result would be applied to shell-type SNRs in order to qualitatively understand their morphology 

Strictly speaking, the Eikonal equation with linear wave treatment is not applicable to waves of finite amplitude, but such waves do not exist in nature, and this is true for any linear theory of astrophysics.
Nevertheless, linear theories are widely used to understand essential aspects of phenomena.
Here we discuss the possibility of applying the MHD wave method to more general SNR cases.
We consider whether shock waves are subject to lensing effects, or reflection and refraction by an inhomogeneous ISM.   

There have been extensive studies of the interaction of expanding shells of SNRs with interstellar clouds \citep{1975ApJ...195..715M,1992ApJ...390L..17S,2012ApJ...744...71I,2021ApJS..253...17S} 
in the premise that the interaction may trigger star formation and acceleration of cosmic rays by the compression of the magnetized molecular gas. 
There have been also a number of hydrodynamical and MHD numerical simulations of the interaction  with the clouds 
\citep{1994ApJ...433..757M,1994ApJ...420..213K,
1990A&A...231..481B,2006ApJS..164..477N,2002AJ....124.2145V,
2003ApJ...584..284V,
2006MNRAS.371..369S,
2014MNRAS.445.2484F,
2014MNRAS.442..229T,2017MNRAS.472.2117M}. 

Commonly seen in these simulations is the implosion of the shock front converging toward the cloud's center axis by refraction of propagation direction according to the deceleration of shock velocity due to momentum conservation.
As a result, the shock wave suffers from a similar lensing effect to the MHD-wave focusing around the low-\va cloud's center, as shown in figures \ref{Vcloud} (A) and \ref{s147up}. 
On the other hand, no numerical simulation has been obtained yet about shock waves interacting with a gaseous cavity or a high-\va cloud containing compressed frozen-in magnetic fields. 
However, we may speculate that a shock wave encountering such a cavity would behave similarly to MHD-waves that expand inside the cavity and make a hole in the front surrounded by a focal ring, as shown in figure \ref{Vcloud} (B).
Also, a shock wave inside high-\va cloud would become sub-Alfv\'enic due to the compression of frozen-in fields, which will propagate as an MHD-wave, so that the behavior would become more similar to that of MHD waves. 

\begin{figure*}   
	\begin{center} 
 (a)
\includegraphics[width=3.3cm]{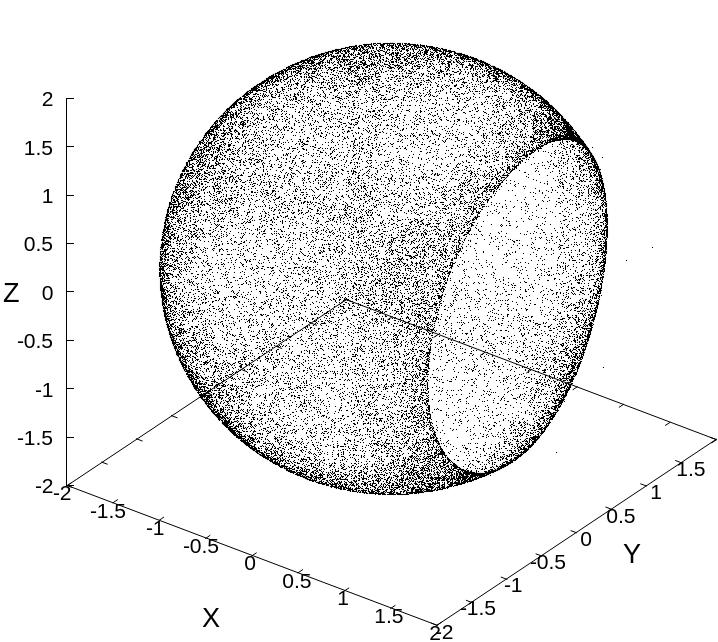} 
\includegraphics[width=3.3cm]{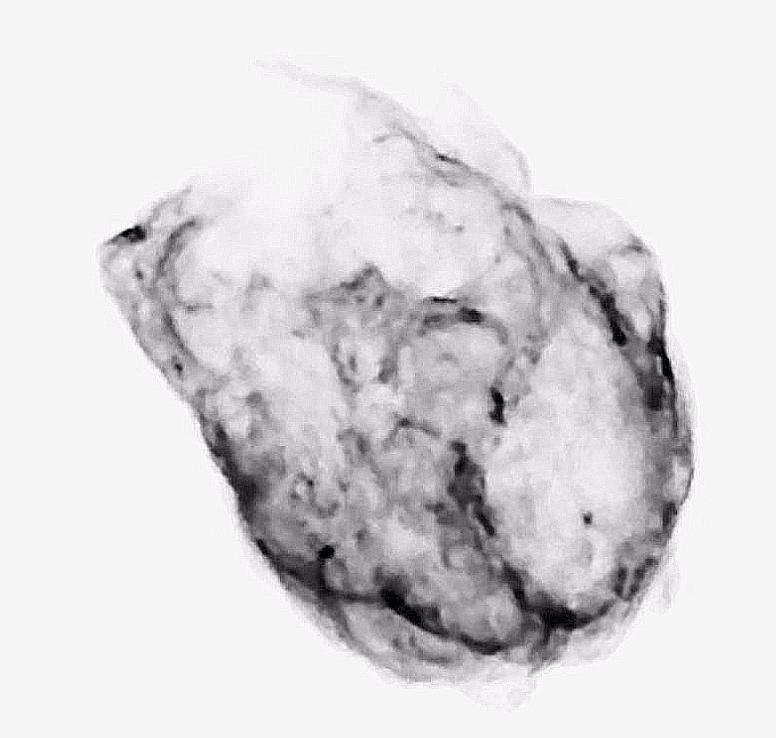}
\hskip 1.5cm
(b) 
\includegraphics[width=3.3cm]{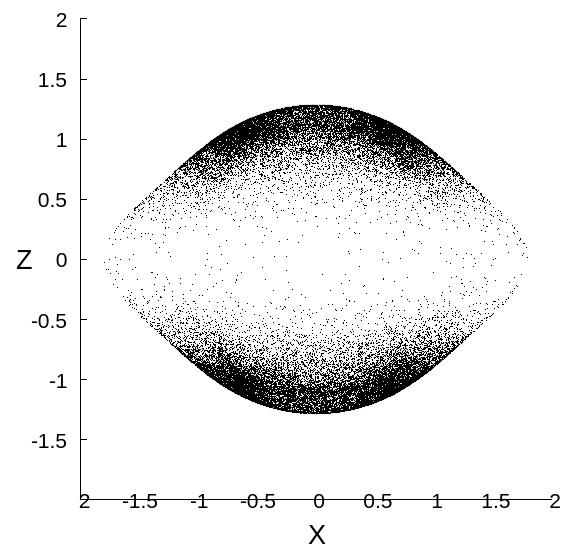} 
\includegraphics[width=3.3cm]{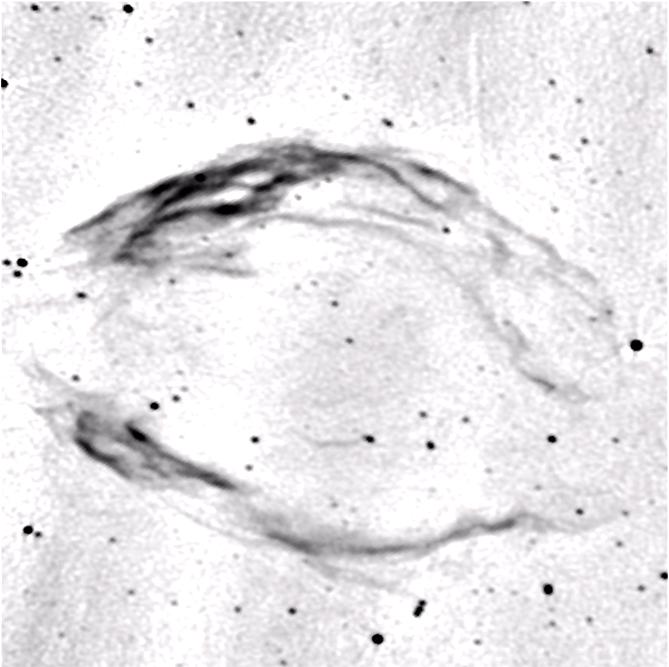}     
\\
\vskip 5mm
(c)
\includegraphics[width=2.3cm]{figs/507-corrugation-xz.jpg} 
\includegraphics[width=2.3cm]{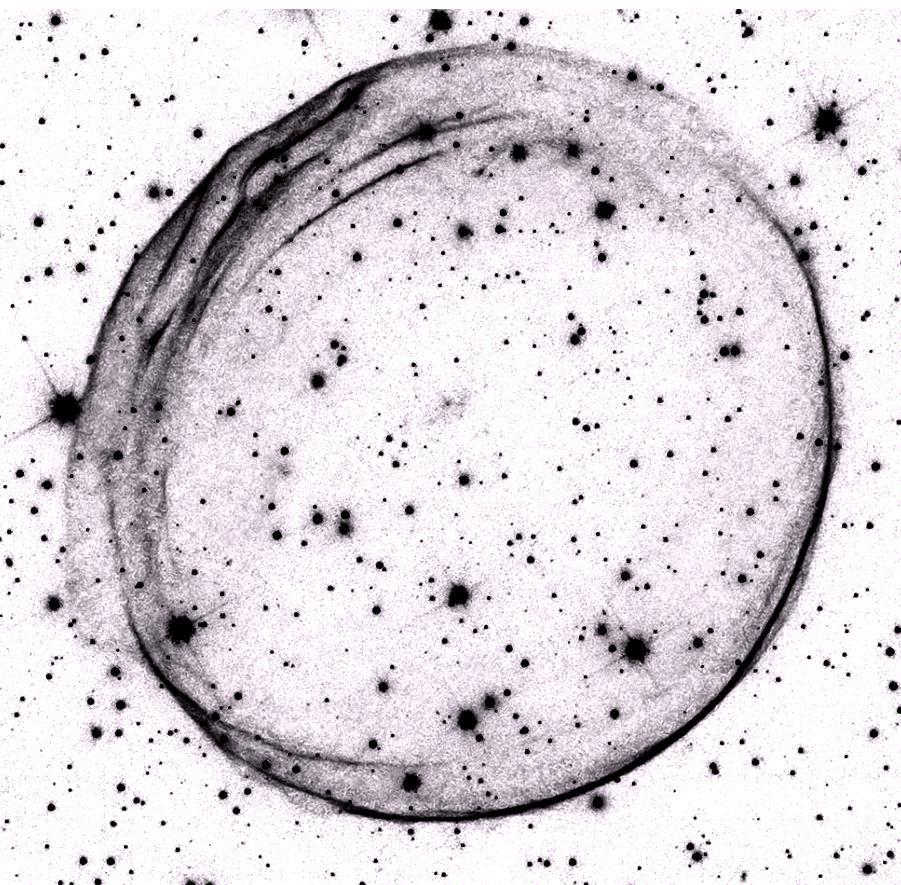} 
\includegraphics[width=2.3cm]{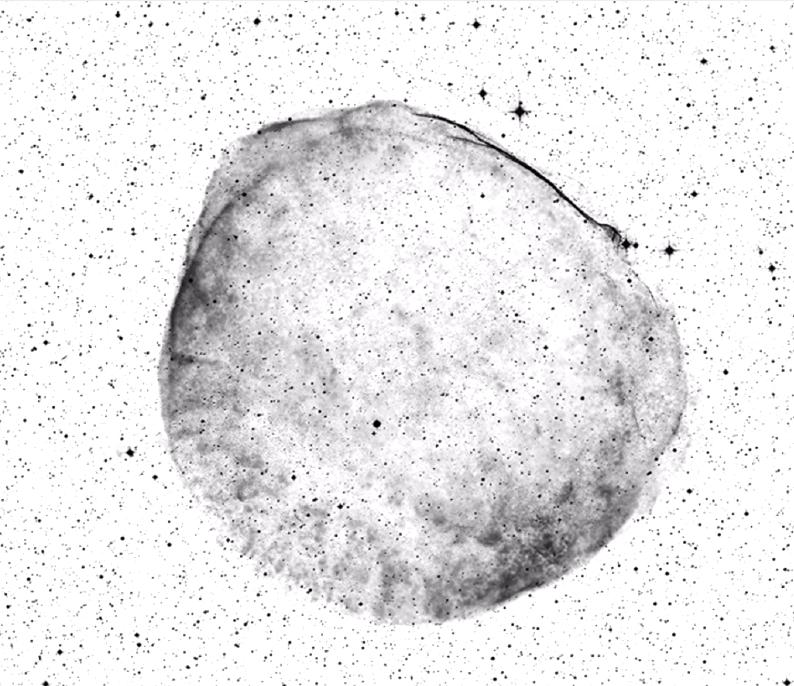} 
\hskip 1.5cm
(d)
\includegraphics[width=2.3cm]{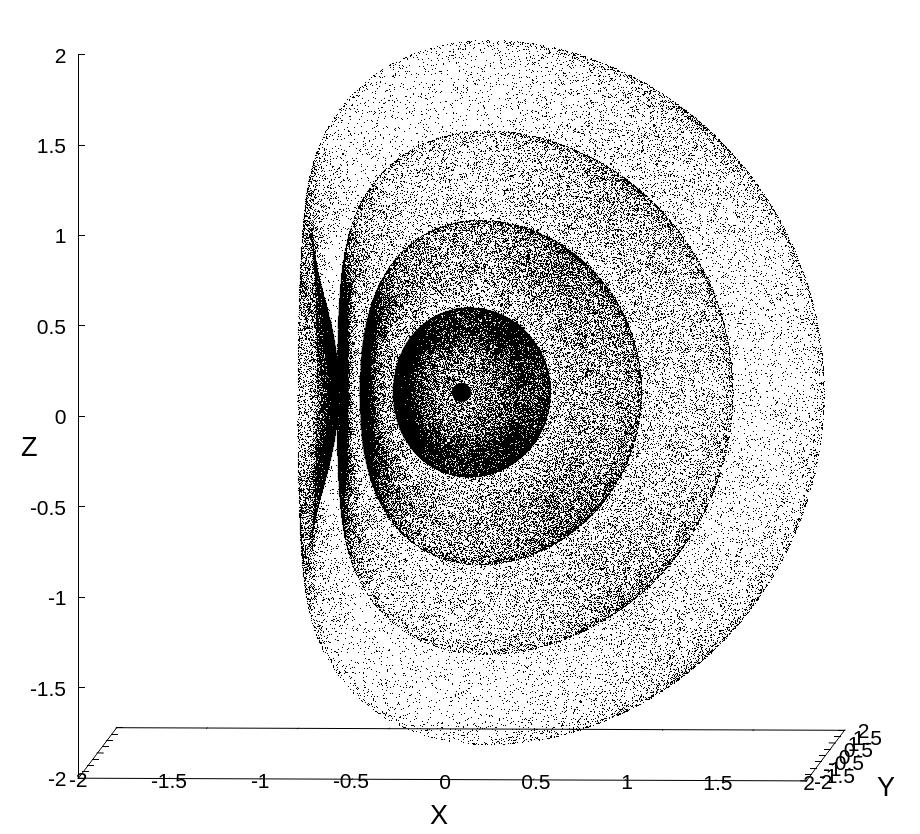}  
\includegraphics[width=2.3cm]{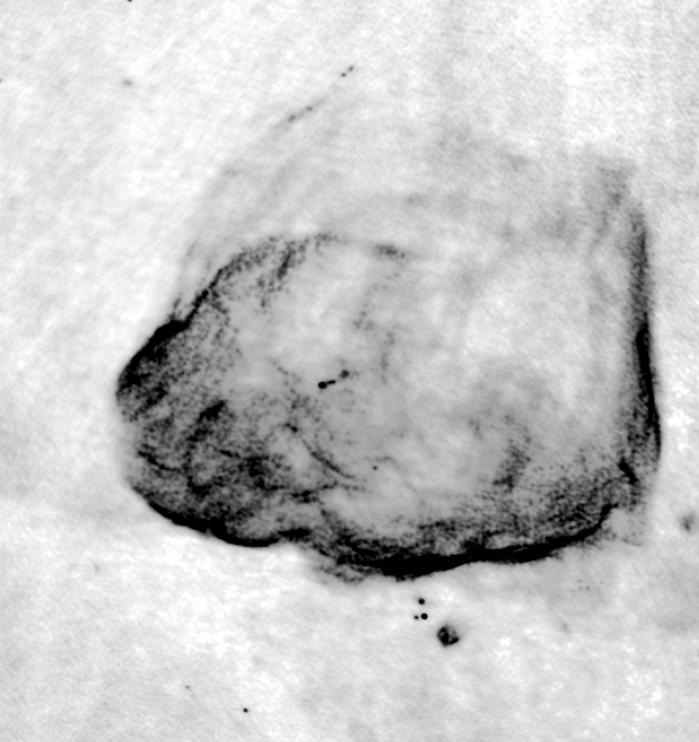} 
\includegraphics[width=2.3cm]{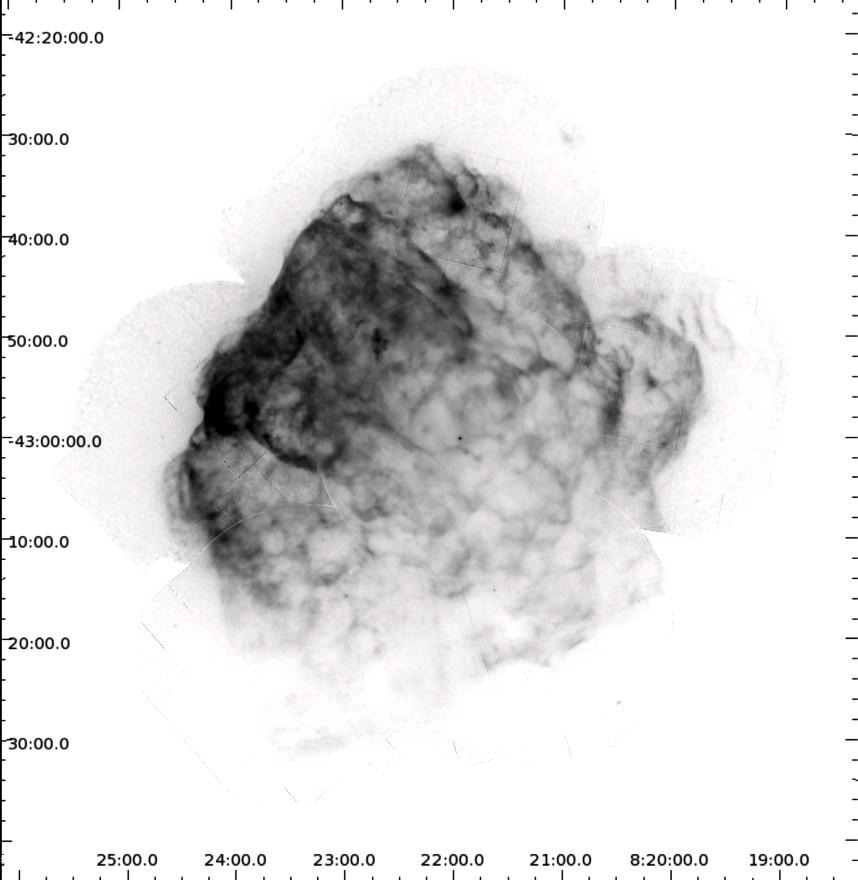}  
\\
\vskip 5mm
(e)   
\includegraphics[width=3.3cm]{figs/514-sinYsinZ.jpg} 
\includegraphics[width=3.3cm]{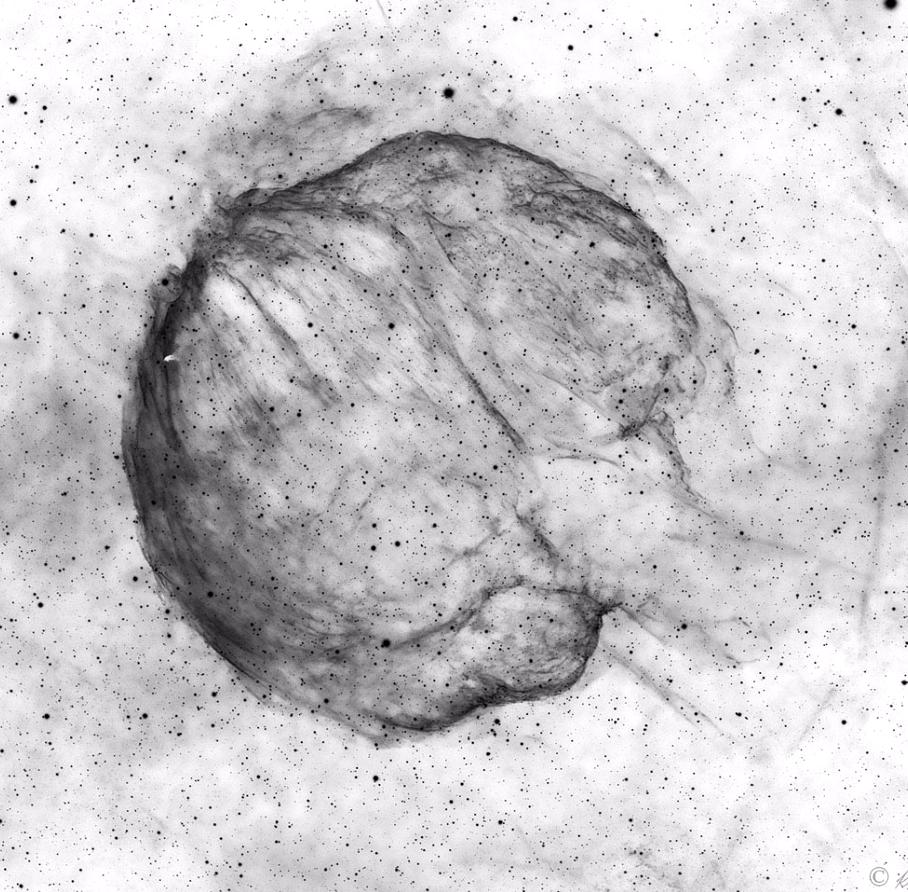}   
\hskip 1.5cm
(f)
\includegraphics[width=3.3cm]{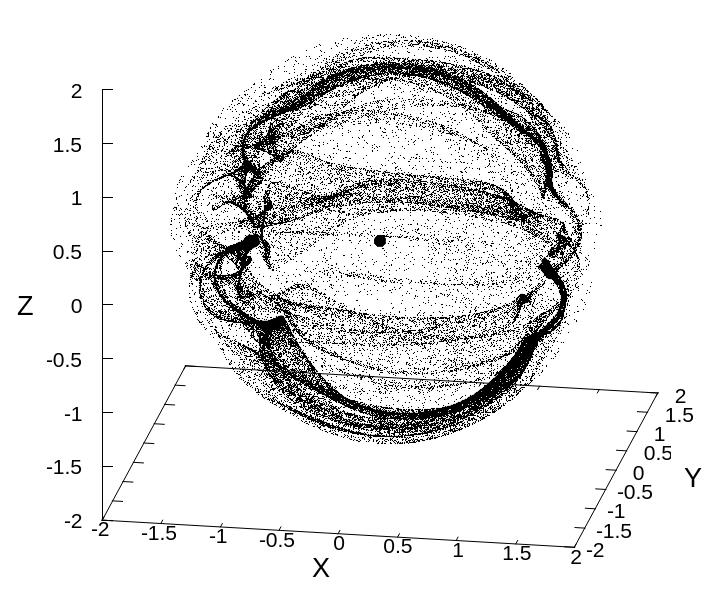}   
\includegraphics[width=3.3cm]{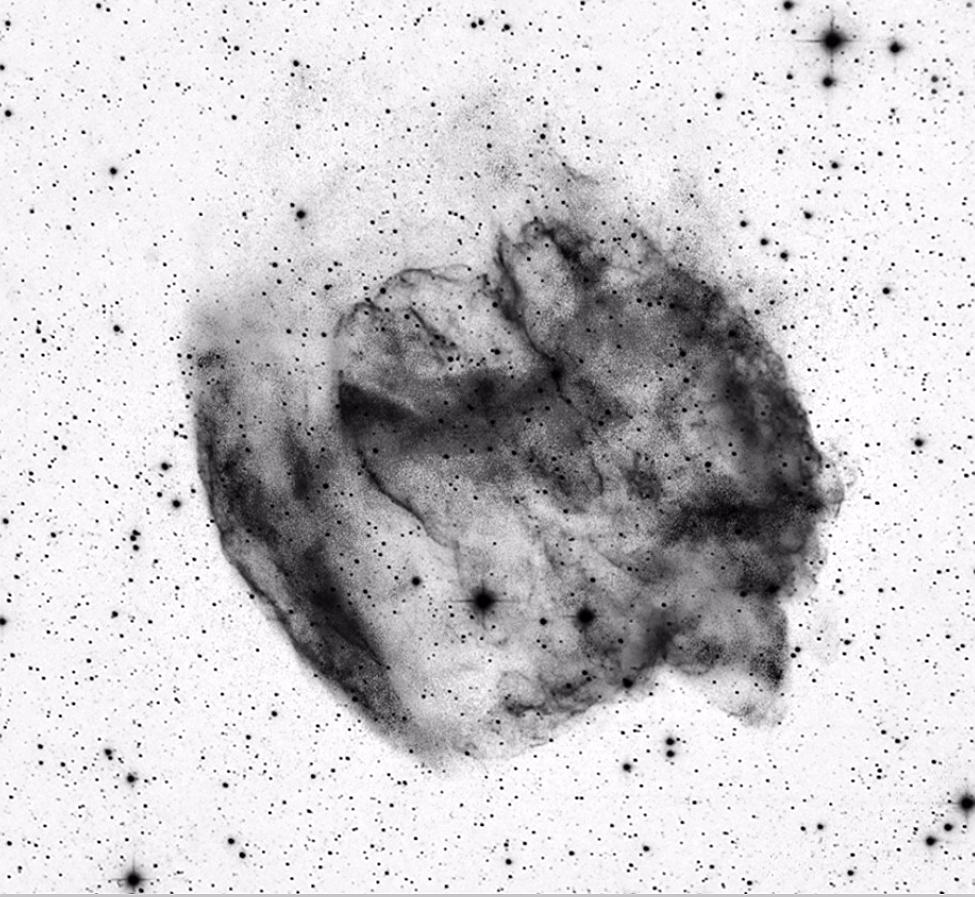}     
\\
\vskip 5mm
(g)   
\includegraphics[width=4cm]{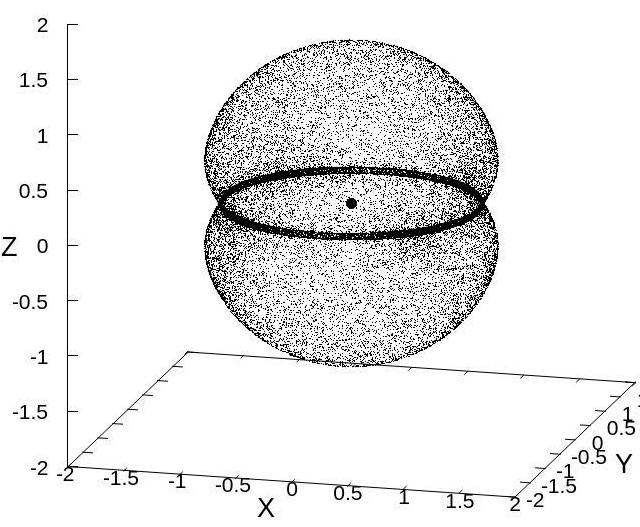}  
\includegraphics[width=3.5cm,angle=90]{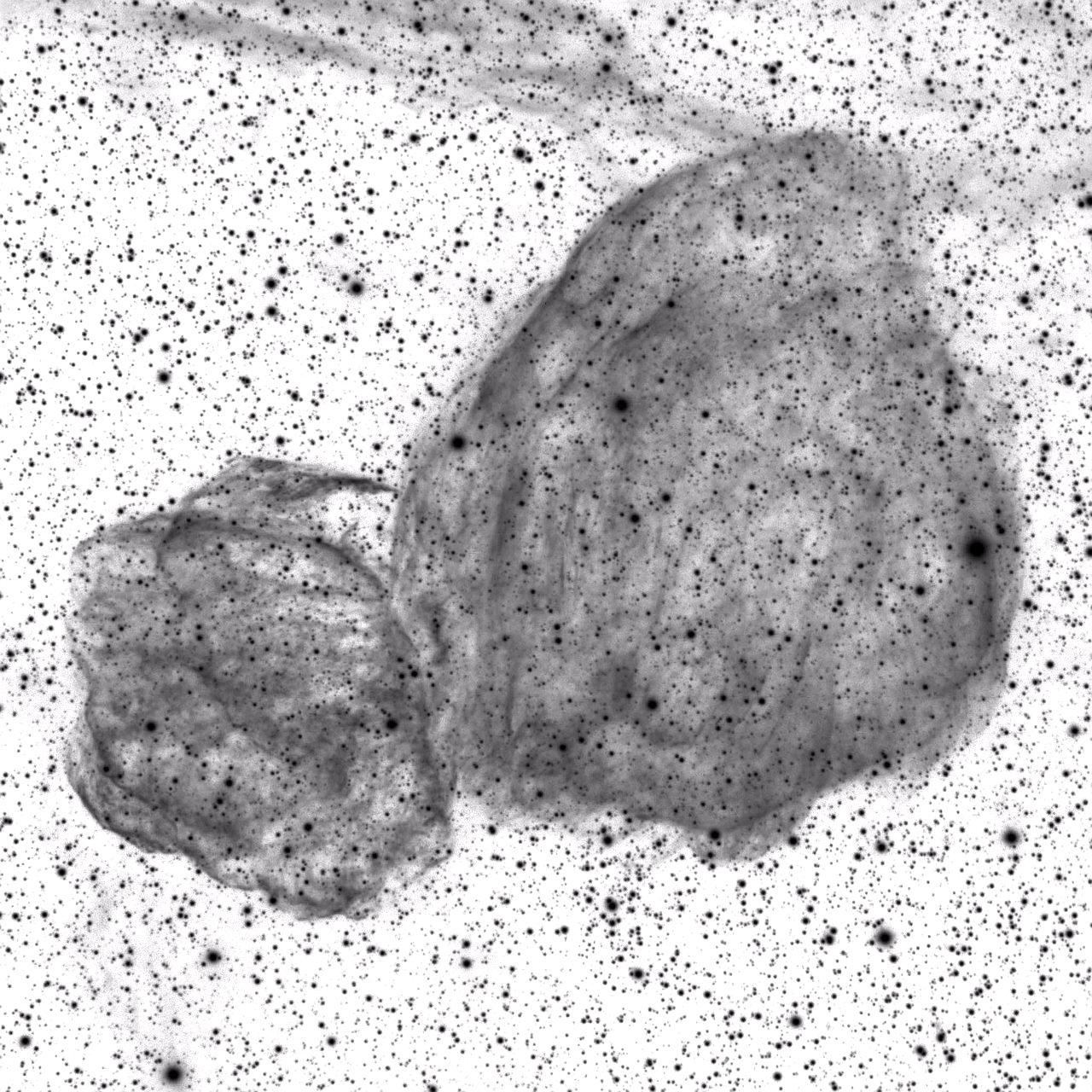}  
\end{center} 
\caption{Gallery of some more examples of peculiar morphology of SNRs possibly influenced by the lensing effect by the inhomogeneity of ISM as compared with the MHD-wave simulation. 
In each block, the left and right panels show the simulation and observed image(s), respectively.
(a) SNR N132-d from the Hubble web site and https://apod.nasa.gov/apod/ compared with the MHD model encountering a high-\va cloud.
(b) Bilateral axisymmetric SNR G296.5+10.0 in radio \citep{2000AJ....119..281G} compaed with MHD model of explosion in a high-\va plate or disc.
(c) SNR 0509-67.5 from the Hubble site (https://stsci-opo.org/STScI-01EVT1GHC7N7YEC39WH79DAN1S.jpg) and SN 1006 (https://hubblesite.org/contents/media/images/) compared with MHD model with slight corrugation on the shell.
(d) SNR G18.8+0.38 from VGPS survey \citep{2006AJ....132.1158S} and Puppis A \citep{2013A&A...555A...9D} compared with MHD model encountering a low-\va plate.
(e) SNR CTB 1 from Hubble web site, NASA, compared with MHD model with sinusoidal variation of \va.
(f) SNR W49b from Hubble web site compared with MHD model with high- and low-\va filaments.
(g) SNR L136 from Hubble web site (https://apod.nasa.gov/apod/image/0801/) compared with MHD model divided by low-\va plate.
}
\label{fig-snrs} 
	\end{figure*}

In summary, shock waves in SNRs behave similarly to MHD waves and result in similar morphological evolution.
The difference is that shock waves disturb/destroy the clouds they pass through, while MHD waves are passive and do not disturb the environment.
The deformation of the shock front may be weaker than that of the MHD wave front, but the sense of deformation is similar to each other.
We therefore believe that the morphological peculiarities observed in shell-type SNRs can be qualitatively explained by the lensing effect of the inhomogeneous ISM.
Figure \ref{fig-snrs} shows some examples of such SNRs and the corresponding MHD wave models.

Through the simulation, we learnt that the shape of MHD wave front suffers from significant deformation due to the lensing effect by the inhomogeneous medium.
This means in turn that a simplified analysis to find association of shell-shaped
molecular gas around a round SNR edge, as often obtained about molecular-gas association \citep{2021ApJS..253...17S}, may not be sufficient to study the past interaction with clouds. 

\ss{Comment on the 'mixed morphology'}

The term, 'mixed morphology' SNR, is often used to represent an SNR with different morphologies observed at different wavelengths such as in radio and X rays, which originate from the outer shell and interior hot gas, respectively \citep{1998ApJ...503L.167R}.
Since the physical properties of the two components are quite different, the difference in their distributions is trivial.
That means, the term may better be read as 'mixed physics', because the existence of such multiple conditions with different spatial distribution is trivial as the general property of a shock-wave driven shell by point explosion.

On the other hand, we have shown that even the radio (or optical) shell, which is the outermost surface of the SNR in contact with the ISM, shows partially different morphologies within the same SNR.
Such "mixed morphologies" with the same physical condition within a single SNR shell are explained by interactions with interstellar clouds with different \alf velocities.

\ss{Different kinds of 'filament'}

We have thus far shown that the filaments in SNRs are classified into two categories, either strings or sheets.
\begin{itemize}
    \item {\bf (a) Sheet filaments:}  
    Many of the observed 'filaments' are tangential view of corrugated (curved) sheets projected on the sky, but not real strings. 
    The corrugation occurs by the reflection and refraction of the wave front by the lens effect of high- and low-\va interstellar fluctuations acting as concave and convex lenses, respectively.    
    Traditionally, most of the shell-type SNRs have been recognized by their loop shapes of tangential-sheet filaments near their edges.    
\item{\bf (b) String filaments}
      A real string (line)-structured filament is formed in such a case that a low-\va tube or an elongated low-\va cloud acts as a cylindrical lens and the waves are converged onto a focal string (line).
      The straight filaments crossing the Cygnus Loop and the $\tau$ nebula are shown to be filaments of this type.
      It must be noted that the string filament exhibits higher amplification of the wave due to the focusing effect compared to the corrugated-sheet filaments.
      An extreme case of wave amplification of this type is the focusing by a spherical low-\va cloud to the focal point, where the wave exhibits a small and concave spherical front with strongly amplified wave. 
\item{\bf (c) Ring around hole}
  The third type of filament is a focusing ring/loop around a hole on the shell surface produced by the diverging reflection of the wave front by a high-\va cloud.
  The tangled rings and arcs of filaments in the Vela nebula and S147 are of this type, which were produced by the diverging-lens effect through many high-\va clouds.
  \item{\bf (d) Focus of converging front:}
  Another type of filament is a dense cluster of converging waves due to focusing effect by a low-\va cloud. The north-eastern clump of high density condensation of filaments in S147 is a typical case of such focusing filaments.  
\end{itemize}
These types of filaments have different formation mechanisms, and hence, different physical conditions.

Sheet filaments are the tangential view of  wave fronts whose large line-of-sight depth makes the apparent brightness much higher than the surroundings.

On the other hand, the string filaments, as well as the converging clumps by convex lenses, are formed by focusing of the waves onto a line or a point, so that the wave amplitude is increased by a factor proportional to the surface-area ratio between before and after the focusing, often amounting to orders of magnitudes.
Such difference of real amplitudes would result in different emission mechanisms and physical conditions in the filaments. However, discussing such difference quantitatively is beyond the present simulation on the linear wave approximation.
It would be a subject for the future work and observations.

\ss{SNRs with unique morphology}

The 'Manatee' nebula (W50) is an SNR having the most strange morphology.
While efforts have been made to explain its elongated shape by interaction of a round shell and a precessing jet from the central compact object, SS433, \citep{2008MNRAS.387..839Z}, the oblique stripes superposed on the surface have been difficult to be understood.
In the present MHD model, however, such unique morphology has been relatively easily reproduced by assuming interaction with the ISM with cigar-shaped high-\va clouds superposed by sinusoidal variation of \va. 

The $\tau$ nebula (N49 in LMC) is also unique for its orthogonal bunches of numerous long and straight filaments \citep{2007AJ....134.2308B}.
This morphology is relatively simply explained in the present MHD model as due to interaction of the SNR surface with two orthogonal bar-shaped clouds having low-\va.
The central long filaments across the Cygnus Loop is classified into the same category due to interaction with an elongated cloud having low-\va.

\section{Summary}
\label{secsummary}

By the simulation we showed qualitatively that most of the morphologies observed in shell-type SNRs can be reasonably explained by the lensing effect due to high- and low-\va interstellar clouds.
Most of the "filaments" in SNRs are shown to be tangential views of corrugated sheets of the wave fronts.
Real string-shaped filaments are formed by cylindrical focusing by cigar-like low \va clouds, and arcs and rings around holes are formed by reflection around high-\va clouds.
The various types of "filaments" on SNR surfaces are, thus, understood as the manifestation of ISM fluctuations, through which the fronts have propagated, suffering from the lensing effect.
{However, since the front is treated as a linear MHD wave, such non-linear quantities as the gas density, temperature, magnetic strength, cosmic-ray acceleration, and electromagnetic emissions cannot be calculated in the present method.}

The SNR morphology can be used to constrain the ISM conditions such as the wavelength of fluctuations, cloud sizes, direction of elongated clouds, and variation of relative amplitudes of the \va velocity.
However, the observed shape is a result of combination of the reflection and refraction paths of the waves, the filaments are significantly displaced from the original position of the lenses (clouds), which makes it not easy to quantitatively determine the parameters of the clouds.
Deconvolution of the observed SNR filaments back into the real ray paths as well as the distribution of \alf velocities in the ISM would be a subject for the future.

\section*{Acknowledgements} 
The computations were carried out at the Astronomy Data Center of the National Astronomical Observatory of Japan. 

\section*{Data availability} 
The images of the supernova remnants were downloaded mainly from the home page of APOD (Astronomy Picture of the Day) of NASA at
https://apod.nasa.gov/apod/, 
and from the Hubble site at
https://hubblesite.org/images.
Some were taken from the web sites described in the figure captions for the individual images.

\section*{Conflict of interest}
The author has no conflict of interest. 
 

\def\aj{AJ} \def\aap{A\&Ap} \def\apjs{ApJ Suppl.}\def\apj{ApJ}\def\apjl{ApJ.L} \def\mnras{MNRAS} \def\pasj{PASJ}

\end{document}